\documentclass[twocolumn]{aastex61}
\usepackage[english]{babel}
\usepackage[utf8]{inputenc}
\usepackage{amsmath}
\usepackage{amssymb}

\received{November 6, 2017}
\revised{March 5, 2018}
\accepted{March 9, 2018}
\submitjournal{ApJ}

\shorttitle{GAMA: impact of group environment}
\shortauthors{S. Barsanti et al.}

\begin{document}

\title{Galaxy And Mass Assembly (GAMA): impact of the group environment on galaxy star formation}

\author{S. Barsanti}
\affil{Department of Physics and Astronomy, Macquarie University, NSW 2109, Australia}
\email{stefania.jess@gmail.com}
  
\author{M. S. Owers}
\affil{Department of Physics and Astronomy, Macquarie University, NSW 2109, Australia}
\affil{The Australian Astronomical Observatory, 105 Delhi Rd, North Ryde, NSW 2113, Australia}

\author{S. Brough}
\affil{School of Physics, University of New South Wales, NSW 2052, Australia}

\author{L. J. M. Davies}
\affil{International Centre for Radio Astronomy Research (ICRAR), University of Western Australia, 35 Stirling Highway, Crawley, WA 6009, Australia}

\author{S. P. Driver}
\affil{International Centre for Radio Astronomy Research (ICRAR), University of Western Australia, 35 Stirling Highway, Crawley, WA 6009, Australia}
\affil{School of Physics and Astronomy, University of St Andrews, North Haugh, St Andrews KY16 9SS, UK}

\author{M. L. P. Gunawardhana}
\affil{Institute for Computational Cosmology, Department of Physics, Durham University, South Road, Durham DH1 3LE, UK}

\author{B. W. Holwerda}
\affil{University of Leiden, Sterrenwacht Leiden Niels Bohrweg 2, NL-2333 CA Leiden, The Netherlands}

\author{J. Liske}
\affil{Hamburger Sternwarte, Universit\"at Hamburg, Gojenbergsweg 112,D-21029 Hamburg, Germany}

\author{J. Loveday}
\affil{Astronomy Centre, University of Sussex, Falmer, Brighton BN1 9QH, UK}

\author{K. A. Pimbblet}
\affil{E. A. Milne Centre for Astrophysics, University of Hull, Cottingham Road, Kingston Upon Hull HU6 7RX, UK}
\affil{School of Physics and Astronomy, Monash University, Clayton, VIC 3800, Australia}
 
\author{A. S. G. Robotham}
\affil{International Centre for Radio Astronomy Research (ICRAR), University of Western Australia, 35 Stirling Highway, Crawley, WA 6009, Australia}

\author{E. N. Taylor}
\affil{Centre for Astrophysics and Supercomputing, Swinburne University of Technology, Hawthorn 3122, Australia}




\begin{abstract}

We explore how the group environment may affect the evolution of star-forming galaxies. We select 1197 Galaxy And Mass Assembly (GAMA) groups at $0.05\leq z \leq 0.2$ and analyze the projected phase space (PPS) diagram, i.e. the galaxy velocity as a function of projected group-centric radius, as a local environmental metric in the low-mass halo regime $10^{12}\leq (M_{200}/M_{\odot})< 10^{14}$. We study the properties of star-forming group galaxies, exploring the correlation of star formation rate (SFR) with radial distance and stellar mass. We find that the fraction of star-forming group members is higher in the PPS regions dominated by recently accreted galaxies, whereas passive galaxies dominate the virialized regions. We observe a small decline in specific SFR of star-forming galaxies towards the group center by a factor $\sim 1.2$ with respect to field galaxies. Similar to cluster studies, we conclude for low-mass halos that star-forming group galaxies represent an infalling population from the field to the halo and show suppressed star formation.
  
\end{abstract}

\keywords{galaxies: evolution -- galaxies: groups: general -- galaxies: kinematics and dynamics -- galaxies: star formation}



\section{Introduction} \label{intro}

The properties of galaxies, such as their star formation rate (SFR), morphology and stellar mass, correlate strongly with the galaxy number density in the surrounding Universe \citep{Dressler1980,Kodama2001,Smith2005,Peng2010}. This correlation is most visible in galaxy clusters, which are the largest halos that have had time to virialize in the Universe, where their cores are found to be dominated by passive galaxies while in their outskirts there is a higher fraction of star-forming galaxies (e.g., \citealp{Balogh1997, Hashimoto1998, Poggianti1999, Couch2001}). The observed correlation with cluster-centric radius reveals radial distance as a crude metric of the time since a particular galaxy has entered the cluster environment - with core galaxies being early virialized cluster members and populations at large radii being increasingly dominated by recently infalling galaxies. However, equating radial distance to the time since infall is a blunt approach as this does not take into account projection effects, and washes out potentially important populations such as, first-pass infalling galaxies which happen to be in the cluster core at the time of observation, backsplash galaxies which have already traversed the cluster core and are observed close to the maximum distance before their second infall, and galaxies which have already undergone multiple passes but appear at large radii. A more sophisticated approach is to classify galaxies based on both position and velocity, considering their dynamical state within the cluster. The projected phase space diagram (PPS), i.e. the galaxy velocity as a function of projected cluster-centric radius, has been extensively used to separate the different cluster populations and to study their spectral features (e.g., \citealp{Pimbblet2006,Mahajan2011,Oman2013,Muzzin2014,Jaffe2015,Oman2016}). These works show that galaxy spectral properties correlate strongly with their position on the PPS. Finally, there is evidence for a relationship between SFR and galaxy density and projected cluster-centric radius, in the sense that star-forming galaxies in clusters show suppressed star formation with respect to field galaxies, and many studies have been dedicated to understanding the underlying physics driving this suppression in galaxy clusters (e.g., \citealp{Lewis2002,Gomez2003,vonderLinden2010,Paccagnella2016}).

The PPS and the role of the environmental mechanisms in affecting galaxy star formation are less clear in low-mass halos, i.e. galaxy groups with mass $\sim 10^{12}-10^{14} M_{\odot}$. Galaxy groups are the most common galaxy environment \citep{Eke2005} and their study offers an important tool for a more complete understanding of galaxy formation and evolution. Similar to cluster environments, several works have found that the galaxy morphology correlates with group-centric distance and local galaxy density for group galaxies (e.g., \citealp{Postman1984, Tran2001, Girardi2003, Brough2006, Wetzel2012}). Moreover, the analysis of massive clusters revealed that the low fraction of star-forming galaxies observed in the dense cluster centers persists in group-like regions beyond the cluster sphere of influence \citep{Lewis2002}. This scenario opens the possibility that galaxies are ``pre-processed'' in groups before they fall into clusters according to a hierarchical scenario of structure formation \citep{Hou2014,Haines2015,Roberts2017}.

Many studies probed the impact of the group environment on star formation in galaxies, spanning a range of epochs (e.g., \citealp{Balogh2011, McGee2011, Hou2013,Mok2014,Davies2016a}). In particular, \citet{Ras2012} and \citet{Ziparo2013} analyzed how the SFR of galaxies in nearby groups depends on radius and local galaxy density. However, they reached conflicting results since \citet{Ras2012} found a decrease by $\sim40$\% of the specific SFR (sSFR=SFR/$M_{*}$) as a function of the projected group-centric distance for star-forming galaxies in 23 nearby galaxy groups ($z\sim0.06$) relative to the field, while \citet{Ziparo2013} observed no decline in SFR and sSFR for the whole galaxy population in 22 groups in the redshift range $0< z < 1.6$. \citet{Wi2012} and \citet{Schaefer2016} both considered group and field galaxies together but obtained opposing conclusions. \citet{Wi2012} showed that the SFR$-$local galaxy density relation is only visible when both the passive and star-forming galaxy populations are considered together, implying that the stellar mass has the largest impact on the current SFR of a galaxy while any environmental effect is not detectable. On the contrary, \citet{Schaefer2016} found that star formation rate gradients in star-forming galaxies are steeper in dense environments with a reduction in total SFR. Finally, the environmental processes responsible for SFR suppression in the halos and the quenching time-scales are still an issue \citep{Wetzel2013, McGee2014, Wetzel2014, Peng2015, Grootes2017}. 

We focus on galaxy groups, considering also clusters to compare the results, and we study the high-fidelity Galaxy And Mass Assembly (GAMA) group catalog since it contains a statistically high number of groups. The aim of this paper is to explore whether and how group environments affect star formation properties of member galaxies. We use the PPS as a proxy for environment and we expand the investigation of the PPS to halos with lower mass $\sim 10^{12}-10^{14} M_{\odot}$ and containing a higher number of galaxies with respect to previous works on groups, in order to probe whether the results found for clusters are seen for lower-mass halos. 

This paper is organized as follows. We present our GAMA group sample, the galaxy member selection and spectral classification in Section~\ref{data}. In Section~\ref{results} we analyze the distributions of passive and star-forming galaxies in radial space, projected phase space and velocity space. We investigate the SFRs of star-forming galaxies as a function of the projected group-centric radius and galaxy stellar mass. Finally, we discuss our results in Section~\ref{discussion} and conclude in Section~\ref{conclusions}. Throughout this work we assume $\Omega_{m}=0.3$, $\Omega_{\Lambda}=0.7$ and $H_{0}=70\hspace{1mm} \rm km\hspace{1mm}s^{-1}\hspace{1mm} Mpc^{-1}$ as cosmological parameters.

\section{Dataset} \label{data}

\subsection{The Galaxy And Mass Assembly survey}

Galaxy And Mass Assembly (GAMA; \citealp{Driver2011,Hopkins2013,Liske2015}) is a spectroscopic and photometric survey of $\sim$300,000 galaxies, down to $r<19.8$ mag and over $\sim$286 degrees$^{2}$ divided in 5 regions called G02, G09, G12, G15 and G23. The redshift range of the GAMA sample is $0<z\lesssim 0.5$ with a median value of $z\sim0.25$. The majority of the spectroscopic data were obtained using the AAOmega multi-object spectrograph at the Anglo-Australian Telescope. GAMA incorporates previous spectroscopic surveys such as the SDSS \citep{York2000}, 2dFGRS \citep{Colless2001,Colless2003}, WiggleZ \citep{Drinkwater2010} and the Millennium Galaxy Catalog (MGC; \citealp{Liske2003,Driver2005}).

The multi-wavelength photometric and spectroscopic data of GAMA cover 21 photometric bands spanning from the far-ultraviolet to the far-infrared, and the spectra cover an observed wavelength range from 3750 to 8850 \AA\hspace{1mm} at a resolution of $R\sim1300$. Considering the combination of the wide area, the high spectroscopic completeness (98.5\% in the equatorial regions; \citealp{Liske2015}), the high spatial resolution and the broad wavelength coverage, the GAMA survey provides a unique tool to investigate the formation and evolution of galaxies in groups.

We use the following already measured optical data: positions and spectroscopic redshifts \citep{Driver2011, Liske2015}, equivalent widths and fluxes of the H$\delta$, H$\beta$, [OIII], H$\alpha$ and [NII] spectral lines \citep{Hopkins2013,Gordon2017}, SFR estimators based on the H$\alpha$ emission lines \citep{Gunawardhana2013} and on the full spectral energy distribution fits \citep{Davies2016,Driver2016}, and stellar masses \citep{Taylor2011}.

\subsection{Group sample}
Our group sample is based on the GAMA Galaxy Group Catalog (G$^{3}$C; \citealp{Robo2011}), built on a Friends-of-Friends (FoF) algorithm which examines both radial and projected comoving distances to assess overlapping galactic halo membership. The radial comoving distances used in the FoF algorithm are derived from the redshifts obtained from the GAMA II data described in \citet{Liske2015}. The group catalog contains 23654 groups (each with $\geq$2 members) and 184081 galaxies from the G09, G12 and G15 regions observed down to $r<19.8$ mag. We select 1197 GAMA galaxy groups by:
\begin{itemize}
\item Group edge: 1.
\item Redshift: $0.05\leq z\leq 0.2$. 
\item Membership: at least 5 members.
\item Mass: $10^{12}\leq (M_{200}/M_{\odot})\leq 10^{15}$.
\end{itemize}

\begin{figure}
\centering
\resizebox{\hsize}{!}{\includegraphics{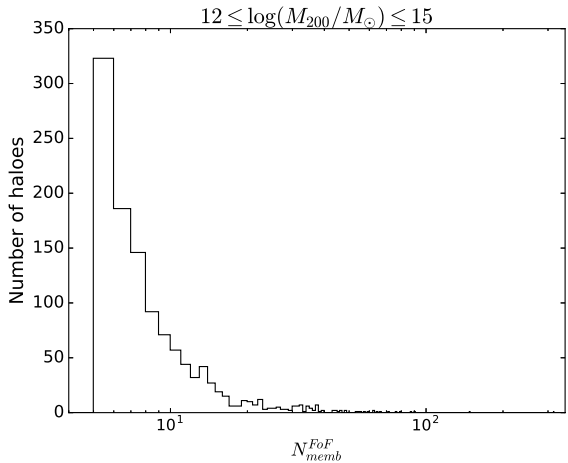}}
\caption{Number of haloes as a function of spectroscopically confirmed galaxies identified by the Friends-of-Friends algorithm ($N^{\rm FoF}_{\rm memb}$). The histogram shows a peak at $N^{\rm FoF}_{\rm memb}=5$.}
\label{histogramNfof}
  \end{figure}

The respective reasons for these chosen criteria are:
\begin{itemize}
\item The group edge quantifies the fraction of the group within the survey volume and group edge=1 means that the group is entirely contained within the survey and we are not just considering a fraction of it.
  \item The minimum $z_{\rm min}=0.05$ is selected in order to minimize the impact of aperture effects due to the 2$''$ fiber used to collect the GAMA galaxy spectra \citep{Kewley2005}. Only 45 groups are present at $0.0\leq z< 0.05$. We choose $z_{\rm max}=0.2$ as the maximum redshift because beyond this the detection of the H$\alpha$ line is unreliable due to the presence of the telluric OH forest at the red end of the spectra. In addition, this allows us to probe low-mass galaxies with stellar mass $M_{*}=10^{9}M_{\odot}$ over the whole redshift range.
\item At least 5 spectroscopically confirmed members identified by the FoF algorithm are needed to obtain reliable estimates of group properties such as velocity dispersion, halo mass and radius \citep{Robo2011}. Figure~\ref{histogramNfof} shows the histogram of member galaxies in the halo sample.
\item As group mass estimator we use $M_{200}$ which is defined as the mass of a spherical halo with a mean density that is 200 times the critical cosmic density at the halo redshift. This study is focused on galaxy groups with $10^{12}\leq (M_{200}/M_{\odot})< 10^{14}$, but we also include clusters with ($M_{200}/M_{\odot})\geq 10^{14}$ in order to compare the results for low-mass halos with those for the high-mass ones. There is no known sharp mass cutoff that divides clusters and groups, but we assume $M_{200}= 10^{14} M_{\odot}$ as a partition mass. $M_{200}$ is estimated using the raw group velocity dispersions calculated by \citet{Robo2011} and according to the $M_{200}-\sigma$ relation of \citet{Munari2013}:
\begin{equation}
 M_{200}=\frac{\sigma^{3}}{1090^{3} h(z)} 10^{15} M_{\odot}   
\end{equation}
where $h(z)=H(z)/(100\hspace{1mm} \rm km\hspace{1mm}s^{-1}\hspace{1mm}Mpc^{-1})$ and $H(z)=H_{0}\sqrt{\Omega_{m}(1+z)^{3}+\Omega_{\Lambda}}$ is the Hubble parameter. We calculate the group radius $R_{200}$ as the radius of a sphere with mass $M_{200}$.
\end{itemize}

\subsection{Selection of member galaxies}
The FoF algorithm tends to assign group membership to galaxies in the very central region of groups, out to $\sim1.5\,R_{200}$. However, we want to investigate the star formation in galaxies out to a projected distance of $9\,R_{200}$, and close to the group redshift. In this Section, we outline our method, which uses both galaxy redshifts and projected distance from the group center, as well as $M_{200}$, to extend the existing spectroscopically confirmed FoF group membership out to $9\,R_{200}$.

The reasons for probing such a large group-centric distance are twofold. First, we would like to compare our results with those of \citet{Ras2012}, who probed out to $\sim10\, R_{200}$. Second, analyzing out to such large radii (i.e. $>4\,R_{200}$) means that we naturally include a benchmark sample of field galaxies that are well beyond the regions where processes related to the group environment are expected to be important. This benchmark sample will be drawn from the same redshift and galaxy stellar mass distributions as the group galaxies, and will therefore allow for an unbiased comparison to be made between the group galaxy properties and those of the benchmark field sample. One potential concern is that of stellar mass-segregation, which may lead to more massive galaxies being preferentially found close to the group center. However, \citet{Kafle2016} showed that there is negligible mass-segregation as a function of radius in the GAMA groups out to 2$\, R_{200}$.

In order to extend our study to 9$\, R_{200}$, probing the group surroundings, we consider also galaxies in the same redshift range, but not assigned to groups by the FoF algorithm. We assign additional galaxies to groups in a manner similar to \citet{Smith2004}, i.e. by minimizing the $C$ parameter as a function of redshift and projected location. Each additional galaxy is assigned only to one group and the $C$ parameter is proportional to the logarithm of the probability that a galaxy is a member of a group assuming that the group velocity distribution is a Gaussian:
\begin{equation}
C=({\rm c}z_{\rm gal}-{\rm c}z_{\rm group})^{2}/\sigma^{2}-4\hspace{1mm}{\rm log}(1-R/R_{\rm group})
\end{equation}
where c is the speed of light, $\sigma$ is the group velocity dispersion, $z_{\rm gal}$ and $z_{\rm group}$ are the redshift of the galaxy and the group respectively, $R$ is the projected distance between the galaxy and the group center, and $R_{\rm group}$ is fixed at 9$\, R_{200}$. The group center is estimated by \citet{Robo2011} as the coordinates of the central galaxy defined with an iterative procedure. 

\begin{figure}
\centering
\resizebox{\hsize}{!}{\includegraphics{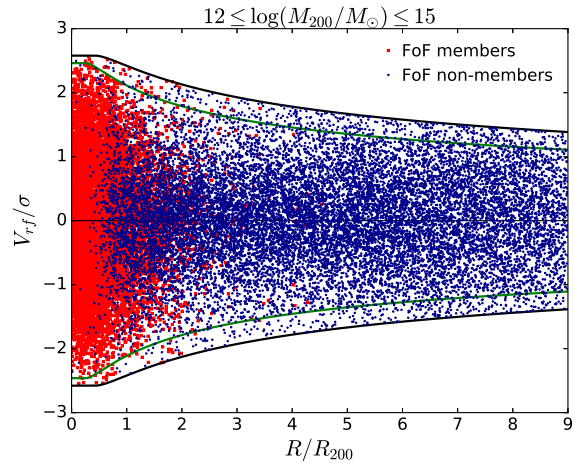}}
\caption{Stacked PPS diagram used to select group/cluster members. Red squares represent FoF members, whereas blue dots are for galaxies not selected by the algorithm but assigned to a halo in order to probe radial distances out to 9$\, R_{200}$. Black and green curves indicate the infall velocities for groups with $M_{200}/M_{\odot}=10^{12}-10^{14}$ and clusters with $M_{200}/M_{\odot}=10^{14}-10^{15}$, respectively.}
\label{PPSmassrange}
\end{figure}

In order to investigate low- and high-mass halos separately and to perform a robust member selection, we define two samples according to their mass: groups with $M_{200}/M_{\odot}=10^{12}-10^{14}$ and clusters with $M_{200}/M_{\odot}=10^{14}-10^{15}$. For each sample we stack both FoF members as well as the galaxies assigned to halos out to 9$\, R_{200}$ and we calculate the infall velocities, i.e. the maximum allowed line-of-sight velocities for group/cluster galaxies. We define only galaxies within these velocities as members. Figure~\ref{PPSmassrange} shows the stacked PPS diagram in normalized units, i.e. $V_{\rm rf}/\sigma$ as a function of $R/R_{200}$ where the galaxy rest-frame velocity is defined as:
\begin{equation}
V_{\rm rf}=\frac{{\rm c}z_{\rm gal}-{\rm c}z_{\rm group}}{(1+z_{\rm group})}.
\end{equation}

The infall velocity, $V_{i}$, is estimated as a function of $x=R/R_{200}$, by considering separately galaxies in each range of halo mass and assuming a Navarro-Frenk-White mass density profile \citep{Navarro1996}:
\begin{equation}
V_{i}(x)=\sqrt{2} V_{c}(x)
\end{equation}
where $V_{c}$ is the circular velocity scaled by $V_{200}=(GM_{200}/R_{200})^{1/2}$ and given by:
\begin{equation}
\left(\frac{V_{c}(R)}{V_{200}}\right)^{2}=\frac{1}{x}\, \frac{\ln(1+\kappa x)-(\kappa x)/(1+\kappa x)}{\ln(1+\kappa)-\kappa/(1+\kappa)}.
\end{equation}
The concentration parameter $\kappa$ is estimated according to the relation of \citet{Dolag2004} and depends on the median $\overline{z}$ and $\overline{M_{200}}$ of the group sample:
\begin{equation}
\kappa(\overline{M_{200}},\overline{z})=\frac{\kappa_{0}}{1+\overline{z}}\left(\frac{\overline{M_{200}}}{10^{14}\,h^{-1}\,M_{\odot}}\right)^{\alpha}
\end{equation}
where $\kappa_{0}=9.59$ and $\alpha=-0.102$ for our cosmological model. The relationship between the concentration and the halo mass justifies our choice to determine the infall velocities for the two samples with different $M_{200}$ ranges: we use $\overline{z}=0.14$, $\overline{M_{200}}=1.5\times 10^{13} M_{\odot}$ and $\overline{z}=0.17$, $\overline{M_{200}}=1.6\times 10^{14} M_{\odot}$ for groups and clusters, respectively. In Figure~\ref{PPSmassrange} the different curves indicate the infall velocities for groups and clusters.

\begin{table}
	\caption{Samples of galaxy groups and clusters.}
	\label{samplegruppi}
        \centering
	\begin{tabular}{l r r r r r}
	  \hline
          \noalign{\smallskip}
	  \hline
          \noalign{\smallskip}
                \multicolumn{1}{l}{$\rm Halos$}
		&\multicolumn{1}{c}{$(M_{200}/M_{\odot})$}
		&\multicolumn{1}{c}{$N_{\rm halos}$} 
		&\multicolumn{1}{c}{$N^{\rm FoF}_{\rm memb}$} 
		&\multicolumn{1}{c}{$N^{\rm FoF}_{\rm non-memb}$}
		&\multicolumn{1}{c}{$N_{\rm tot}$}
		\\
		\hline
                \noalign{\smallskip}
                Groups & $10^{12}-10^{14}$ & 1104 & 10027 & 11762 &21789\\
		Clusters &$10^{14}-10^{15}$ &   93 & 2774  & 5200  & 7974\\
                \noalign{\smallskip}
                \hline
                 \noalign{\smallskip}
	\end{tabular}	
\end{table}

Table~\ref{samplegruppi} lists the halo mass range of each sample ($M_{200}/M_{\odot}$), the number of halos ($N_{\rm halos}$), the number of members identified by the FoF algorithm ($N^{\rm FoF}_{\rm memb}$), the number of galaxies not selected by the algorithm but assigned to a halo ($N^{\rm FoF}_{\rm non-memb}$) and the resulting total number of members ($N_{\rm tot}$).

\begin{figure}[h!]
\centering
\resizebox{\hsize}{!}{\includegraphics{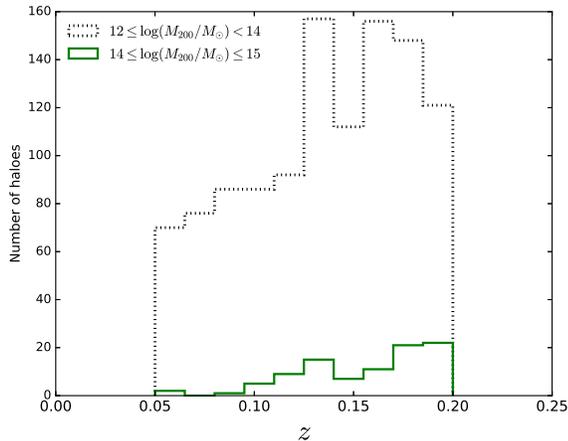}}
\resizebox{\hsize}{!}{\includegraphics{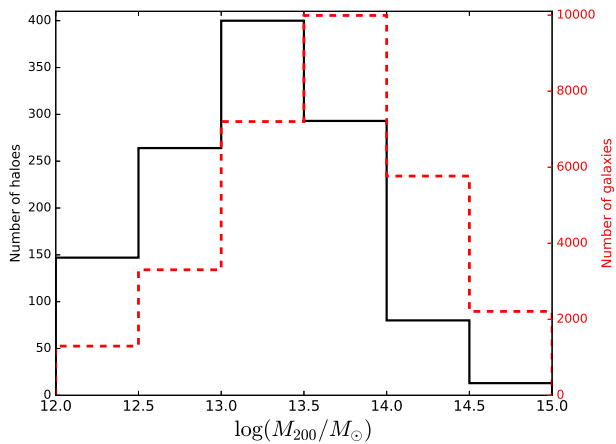}}
\caption{\textit{Upper panel}: Number of groups and clusters as a function of $z$ (dotted black and solid green lines, respectively). \textit{Lower panel}: Number of halos and member galaxies as a function of $M_{200}$ (solid black and dashed red lines, respectively). Most halos have $10^{13}\leq (M_{200}/M_{\odot})\leq 10^{14}$ and the majority of member galaxies belong to these groups.}
\label{histogramsZM200}
\end{figure}

Figure~\ref{histogramsZM200} shows the number of halos as a function of redshift and halo mass. Groups and clusters show peaks at higher redshift because in that range a larger volume of targets has been probed. Most halos have $10^{13}\leq (M_{200}/M_{\odot})\leq 10^{14}$ and the majority of member galaxies belong to these groups. In this context, this study represents a further step with respect to that of \citet{Oman2016}, as well as of \citet{vonderLinden2010}, since both of these works contain low-mass halos with masses $<10^{14} M_{\odot}$, but their satellite numbers are dominated by galaxies in clusters with mass $\geq 10^{14} M_{\odot}$, while we are probing the group mass regime with the majority of galaxies. 

\subsection{Spectroscopic classification of galaxies}
Our primary aim in this paper is to investigate the properties of star-forming galaxies in groups. We select member galaxies with stellar mass $10^{9}\leq (M_{*}/M_{\odot})\leq 10^{12}$ to include both low- and high-mass galaxies and we identify the passive and star-forming populations. We consider only galaxies with a signal-to-noise ratio (S/N) per pixel greater than 3 in the 6383$-$6536 \AA\hspace{1mm}window.

Our spectroscopic classification is outlined below. First, we divide galaxies into two broad categories according to the features of their spectra: absorption- and emission-line galaxies. The absorption-line spectra are typical of galaxies dominated by an old and evolved stellar population, while emission-line galaxies can be characterized by young and new stars, by an AGN or by a Low-Ionization Nuclear Emission-line Region that have low levels of star formation activity and old stars. We use the convention of positive equivalent widths (EW) for emission lines and a negative sign for the absorption features. Since the H$\alpha$ emission line is an excellent tracer of star formation, we use it to select emission-line galaxies. We follow \citet{CidFernandes2011} and define emission-line galaxies as those having $\rm EW(H\alpha)>3$\AA, and add the additional criteria that the S/N of the [NII] line must be larger than 3. The latter criterion helps to guard against spurious single line detections. Absorption-line galaxies are all the remaining ones with $\rm EW(H\alpha)\leq 3$\AA\hspace{1mm}(see Table~\ref{EAclass}).

From the absorption-line galaxies we select the passive members with $\rm EW(H\delta)\geq -3$\AA\hspace{1mm}in order to avoid the contamination from H$\delta$ strong and post-starburst galaxies \citep{Goto2007,Paccagnella2017}. However, the H$\delta$ strong and post-starburst galaxies are $<1$\% and their inclusion does not affect our results. For the emission-line galaxies a further classification is needed to determine whether the emission is due to star formation, since the $\rm EW(H\alpha)> 3$\AA\hspace{1mm}cut does not necessarily imply that a galaxy is star-forming, as it includes AGN and composite systems. Thus, the emission-line galaxies with $\rm S/N>3$ of the H$\beta$, [OIII], H$\alpha$ and [NII] lines are classified as star-forming or AGN according to the \citet{Kauffmann2003} prescription, based on the flux ratios [NII]/H$\alpha$ and [OIII]/H$\beta$:
\begin{equation}
  \log(\rm{[OIII]/H\beta})\leq \frac{0.61}{\log(\rm{[NII]/H\alpha})-0.05}+1.3
\end{equation}

\begin{figure}
\centering
\resizebox{\hsize}{!}{\includegraphics{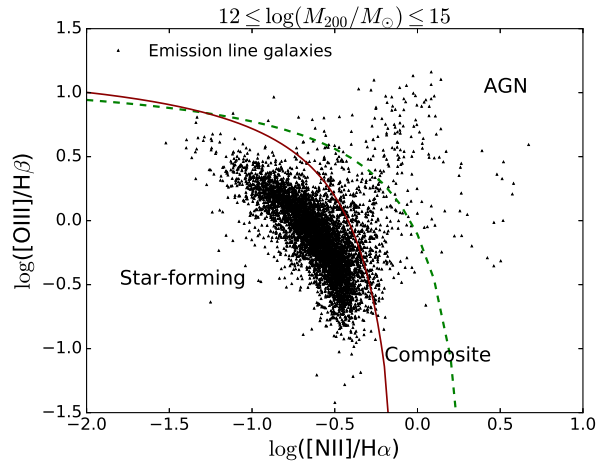}}
\caption{Stacked BPT diagram for emission-line galaxies with S/N$>3$ of the H$\beta$, [OIII], H$\alpha$ and [NII] lines (black triangles) to select star-forming galaxies. The red line represents the adopted star-forming/AGN classification of \citet{Kauffmann2003}. The dashed green line shows the extreme-starburst model defined in \citet{Kewley2001}. Galaxies that fall in the region between the Kewley and Kauffmann lines are classified as composite galaxies.}
\label{BPT}
\end{figure}

We choose the \citet{Kauffmann2003} classification in order to avoid contamination by composite galaxies. Prior to measuring the line ratios, the fluxes of the Balmer lines are corrected for the underlying stellar absorption in the following way \citep{Hopkins2003}:
\begin{equation}
 \rm F_{H\lambda}=\left(\frac{EW_{H\lambda}+EW_{c}}{EW_{H\lambda}}\right)f_{H\lambda}
\end{equation}
where $\rm f_{H\lambda}$ is the observed H$\lambda$ flux with $\lambda=\alpha$ or $\beta$, and $\rm EW_{c}=2.5$\AA\hspace{1mm}is the constant correction factor \citep{Hopkins2013,Gordon2017}. Figure~\ref{BPT} displays the BPT \citep*{Baldwin1981} diagram, i.e. log([OIII]/H$\beta$) versus log([NII]/H$\alpha$), used to identify star-forming galaxies and AGNs. There are 7990 galaxies where either H$\alpha$ or H$\beta$ or [OIII] had $\rm S/N<3$. For those cases, we follow \citet{CidFernandes2011} and define star-forming galaxies as those with log([NII]/H$\alpha$)$\leq -0.4$.

Table~\ref{EAclass} summarizes the constraints for the spectroscopic classification of member galaxies (Col. from 2 to 5) and reports their numbers (Col. 6).

\begin{table*}[t]
  \caption{Spectroscopic classification of galaxies.}
  \label{EAclass}
  \centering
\begin{tabular}{l c c c c r}
  \hline
  \noalign{\smallskip}
  \hline
  \noalign{\smallskip}
\multicolumn{1}{l}{\small{Spectral type}}
&\multicolumn{1}{c}{\small{EW H$\alpha$}}
&\multicolumn{1}{c}{\small{EW H$\delta$}}
&\multicolumn{1}{c}{\small{log([NII]/H$\alpha$)}}
&\multicolumn{1}{c}{\small{S/N [NII]}}
&\multicolumn{1}{c}{\small{$N_{\rm gals}$}}
\\
\multicolumn{1}{l}{}
&\multicolumn{1}{c}{\small{\AA}}
&\multicolumn{1}{c}{\small{\AA}}
&\multicolumn{1}{c}{}
&\multicolumn{1}{c}{}
&\multicolumn{1}{c}{\small{}}
\\
\hline
\noalign{\smallskip}
\small{Absorption}   & $\leq 3$ & &  &  & 12000\\
\small{Emission}  & $>3$  & & & $>3$ & 13500\\
\noalign{\smallskip}
\hline
\noalign{\smallskip}
\small{Passive}   &$\leq 3$   & $\geq -3$ & & &  10663\\
\small{Star-forming}  & $>3$  & & $\leq -0.4$ & &  10239\\
\small{AGN/Composite}  &$>3$ & & $> -0.4$ & &  3261\\
\noalign{\smallskip}
\hline
\noalign{\smallskip}
\end{tabular}
\tablecomments{There are 3 galaxies without measured EW(H$\alpha$) and 3 without measured EW(H$\delta$).}
\end{table*}

\subsection{SFR estimators}
In this work we analyze whether and how the star formation activity is affected by the group/cluster environment with respect to the field. We use two different estimators of SFR taking advantage of the spectroscopic and photometric GAMA data. The spectroscopic estimator only probes the emission lines in the $2''$ aperture of the fiber and measures an ``instantaneous'' SFR, as it probes star formation from the last $\sim$10 Myrs. The photometric SFR measurement includes light from the whole galaxy and is averaged over a longer timescale. Therefore, the two probes are complementary (see \citealp{Davies2016} for the details on scaling relations).

The spectroscopic SFR ($\rm SFR_{H\alpha}$) is calculated assuming a \citet{Salpeter1955} initial mass function:

\begin{equation}
 \rm{SFR_{ H\alpha}}=\frac{L_{\rm{H}\alpha}}{1.27\times 10^{34}}
\end{equation}

where the H$\alpha$ luminosity (L$_{\rm H\alpha}$) is estimated following the procedure outlined in \citet{Hopkins2003} and \citet{Gunawardhana2013}. Briefly, the galaxy's $r$-band magnitude is used in combination with the fibre-based EW(H$\alpha$) to determine an approximately aperture-corrected, total L$_{\rm H\alpha}$. A constant 2.5\AA\hspace{1mm}is added to EW(H$\alpha$) to account for stellar-absorption, and the Balmer decrement is used to correct for dust obscuration (see \citealp{Gunawardhana2011,Gunawardhana2013} for a detailed explanation).

The photometric SFR ($\rm SFR_{MAGPHYS}$) is obtained with the spectral energy distribution (SED)-fitting code MAGPHYS \citep{daCunha2008,Davies2016,Driver2016}, which compares models of ultraviolet/optical/near-infrared spectral templates of stellar populations and mid-/far-infrared templates of dust emission with the observed photometry of a given galaxy, determining the overall best fit stellar+dust template. The MAGPHYS code provides an estimate of the galaxy SFR averaged over the last 100 Myr using a best-fitting energy balance model, where the obscuration-corrected SED is determined by balancing energy absorbed in the ultraviolet/optical with that emitted in the infrared.

\begin{figure}
\centering
\resizebox{\hsize}{!}{\includegraphics{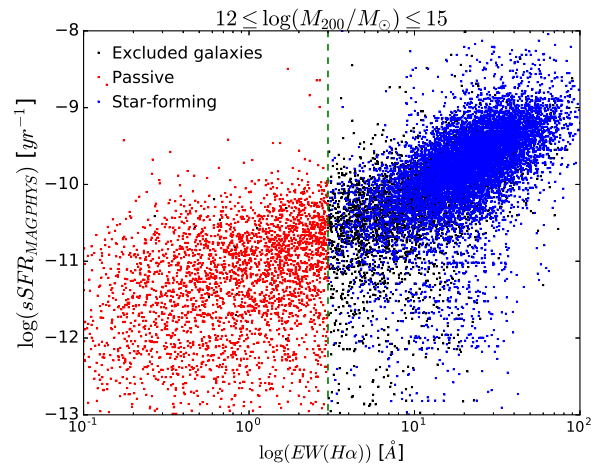}}
\caption{$\rm sSFR_{MAGPHYS}$ as a function of $\rm EW(H\alpha)$ for the whole galaxy population. The dashed green line represents the $\rm EW(H\alpha)>3$\AA\hspace{1mm} cut used to select star-forming galaxies (blue squares) from passive galaxies (red squares). Black squares represent excluded galaxies with $\rm EW(H\alpha)>3$\AA\hspace{1mm} ($\rm EW(H\alpha)\leq3$\AA\hspace{1mm}) and identified as AGN/composite (H$\delta$ strong/post-starburst).}
\label{sSFRvsHalpha}
\end{figure}

Both the spectroscopic and photometric SFR estimators are calculated for the star-forming galaxies spectroscopically selected according to the procedure described in Section 2.4. We show in Figure~\ref{sSFRvsHalpha} how the $\rm EW(H\alpha)>3$\AA\hspace{1mm} cut affects the selection of star-forming galaxies and maps into a limit in sSFR. We use $\rm sSFR_{MAGPHYS}$ since it is not possible to measure $\rm sSFR_{H\alpha}$ for all galaxies.

\section{ANALYSIS AND RESULTS} \label{results}
We explore how the group environment may affect galaxy star-forming properties. In order to pursue our aim we use two different samples of halos and galaxies, i.e. the Full Sample and the Restricted Sample. Their features are described below and are summarized in Table~\ref{SamplesAB}:

\begin{itemize}
\item Full Sample: this includes all the halos at $0.05\leq z\leq 0.2$ and galaxies with $10^{9}\leq (M_{*}/M_{\odot})\leq 10^{12}$ out to 9$\,R_{200}$ (see Sections 2.2$-$2.4).
\item Restricted Sample: this is characterized by smaller ranges in redshift and stellar mass compared to the Full Sample. It includes halos at $0.05\leq z\leq 0.15$ and galaxies with $10^{10}\leq (M_{*}/M_{\odot})\leq 10^{12}$ out to 9$\,R_{200}$. The chosen $z$ and $M_{*}$ limits correspond to a completeness of $\sim 95$\% according to Figure 6 of \citet{Taylor2011} (the grey line refers to our galaxy sample observed down to $r<19.8$ mag), who estimated the GAMA stellar mass completeness limit as a function of redshift.  
\end{itemize}

We perform the following analyses using the Full Sample and the Restricted Sample in the Sections listed in Table~\ref{SamplesAB} for the following reasons:
\begin{itemize}
\item In Sections 3.1 and 3.2 we compare the fractions of passive (PAS) and star-forming (SF) galaxies in radial and projected-phase spaces, respectively. We use the Restricted Sample. The stellar mass cut is necessary because we measure fractions for different galaxy populations, i.e passive fraction and star-forming fraction. At a given $r$-band magnitude, blue star-forming galaxies have a lower stellar mass when compared with red passive galaxies. Therefore, probing a galaxy stellar mass range that is not complete would bias against passive galaxies for a given stellar mass, and affect the measured fractions.
\item In Section 3.3 we investigate the distribution of the passive and star-forming populations in velocity space. We use the Full Sample, without taking into account the stellar mass completeness limit since we are not studying galaxy fractions.
\item In Sections 3.4$-$3.8 we focus only on star-forming galaxies of the Full Sample and study the dependence of star formation rate (SFR) on group-centric radius and galaxy stellar mass.   
\end{itemize}

\begin{table}[h!]
  \caption{Full and Restricted Samples of halos and galaxies.}
  \label{SamplesAB}
  \centering
\begin{tabular}{c c c}
  \hline
  \noalign{\smallskip}
  \hline
  \noalign{\smallskip}
  \multicolumn{1}{l}{}
  &\multicolumn{1}{c}{Full}
&\multicolumn{1}{c}{Restricted}
\\
\hline
\noalign{\smallskip}
$N_{\rm halos}$ & 1197& 679\\
$z$ &  0.05$-$0.20&0.05$-$0.15 \\
$(M_{200}/M_{\odot})$  & $10^{12}-10^{15}$& $10^{12}-10^{15}$\\
$(M_{*}/M_{\odot})$ & $10^{9}-10^{12}$& $10^{10}-10^{12}$\\
\noalign{\smallskip}
\hline
\noalign{\smallskip}
Sections  & 3.3$-$3.6&3.1$-$3.2 \\
\noalign{\smallskip}
\hline
\noalign{\smallskip}
\end{tabular}
\end{table}

\subsection{Fractions of galaxies in radial space}
\begin{figure*}
	\centering
\includegraphics[scale=0.40]{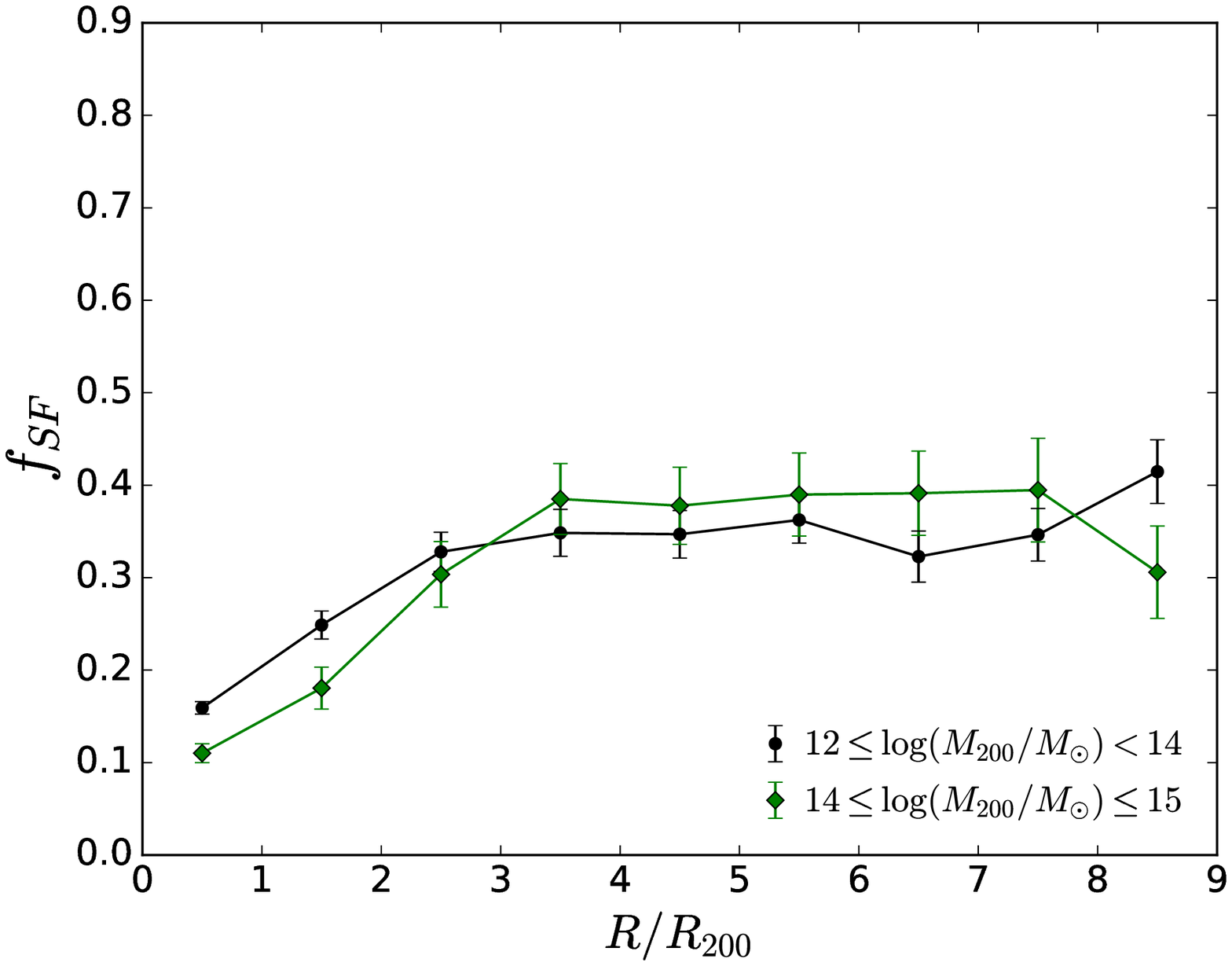}	
\includegraphics[scale=0.40]{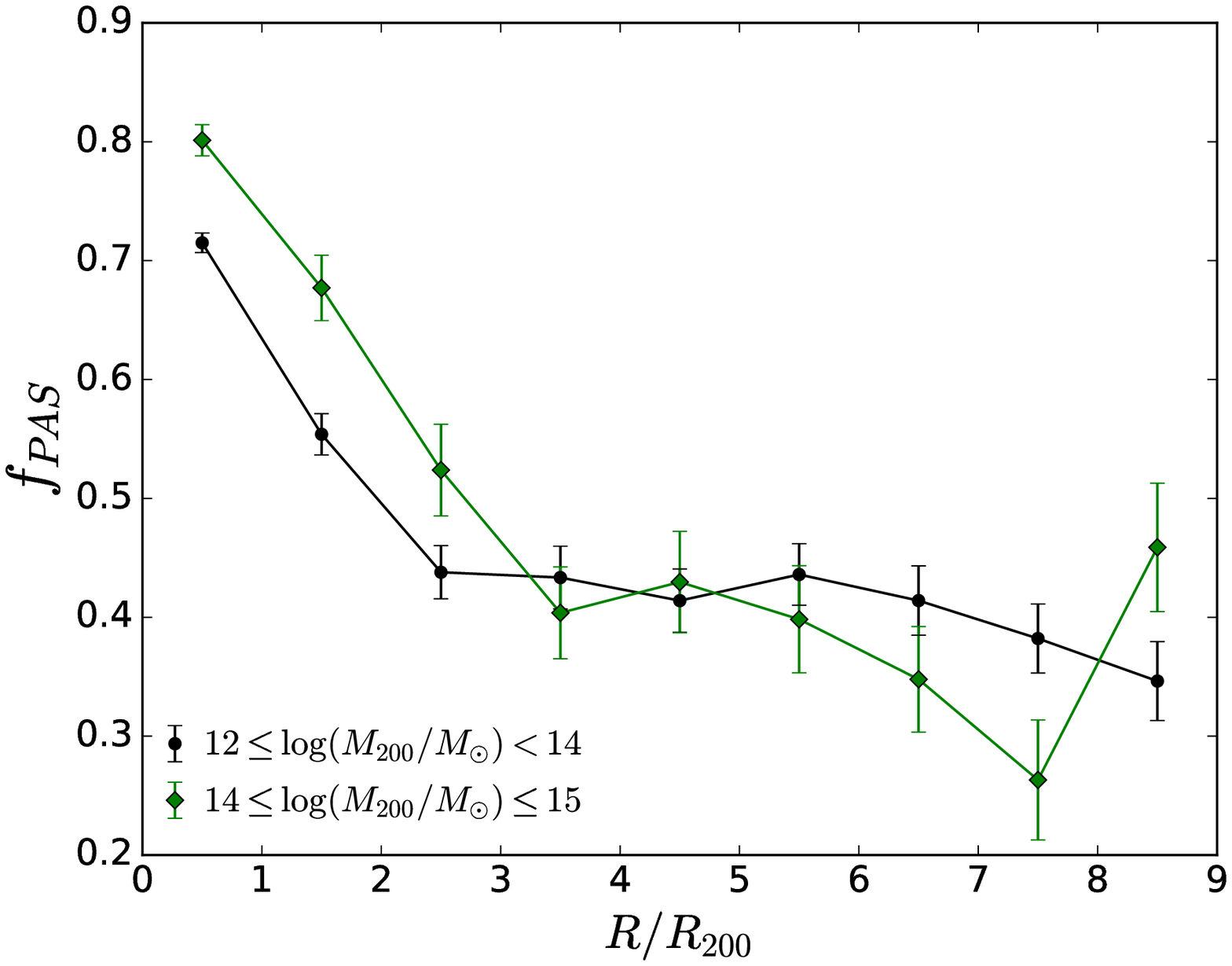}
	\caption{Fractions of star-forming (\textit{left panel}) and passive galaxies (\textit{right panel}) as a function of projected radius for groups (black dots) and clusters (green diamonds). We consider 9 radial bins and binomial error bars. The fraction of passive galaxies strongly decreases from the inner halo regions to large radii, while the fraction of star-forming galaxies increases towards the outskirts.}
	\label{FractionsvsRadius}
\end{figure*}

Different works confirm that group galaxy properties such as morphology, color and spectral type all correlate with group-centric distance  (e.g., \citealp{Carlberg2001,Tran2001, Girardi2003, Wetzel2012, Hou2014}). In order to test the presence of the passive versus star-forming$-$radius relation, we consider the Restricted Sample and explore the fractions of star-forming and passive galaxies as a function of group-centric radius in Figure~\ref{FractionsvsRadius}. The fraction of each galaxy population is estimated with respect to the total galaxy sample containing passive, H$\delta$ strong, post-starburst, star-forming galaxies and AGNs/composites. Table~\ref{tableGalaxyPop} lists the numbers of halos ($N_{\rm halos}$), star-forming ($N_{\rm SF}$) and passive ($N_{\rm PAS}$) galaxies for each $M_{200}$ range.

Figure~\ref{FractionsvsRadius} shows that the fraction of passive galaxies strongly decreases from the inner halo regions to $\sim3.5\,R_{200}$ in clusters and $\sim2.5\,R_{200}$ in groups (\textit{right panel}), while the fraction of star-forming galaxies increases out to the same radii (\textit{left panel}). Beyond these distances both the fractions remain approximately constant. In groups the passive fraction declines by a factor $\sim2$ at 2.5$\,R_{200}$, while the star-forming fraction rises by a factor $\sim1.5$ at the same radius. The maximum is $f_{\rm SF}\sim0.40$ at 9$\,R_{200}$ because of the selected range in stellar mass $10^{10}\leq (M_{*}/M_{\odot})\leq 10^{12}$ and there are fewer star-forming objects with higher mass \citep{Taylor2015}. Our results confirm that the passive versus star-forming$-$radius relation is present in galaxy groups as well as in the more studied cluster environment and that star-forming galaxies are mainly found in the halo outskirts, in agreement with previous works (e.g., \citealp{Whitmore1993,Tran2001, Girardi2003, Goto2003b, Brough2006, Wetzel2012, Hou2014, Fasano2015}).

\begin{table}[h!]
 	\caption{Restricted Sample: galaxy populations out to 9$\,R_{200}$.}
 	\label{tableGalaxyPop}
        \centering
 	\begin{tabular}{l r r r r}
 		\hline
                \noalign{\smallskip}
	        \hline
                \noalign{\smallskip}
 		\multicolumn{1}{l}{$(M_{200}/M_{\odot})$}
                &\multicolumn{1}{c}{$N_{\rm halos}$}
 		&\multicolumn{1}{c}{$N_{\rm SF}$}
 		&\multicolumn{1}{c}{$N_{\rm PAS}$}
 		\\
 		\hline
                \noalign{\smallskip}
                $10^{12}-10^{14}$ &643 & 1486 &3543 \\
 		$10^{14}-10^{15}$ &36 & 465 & 1294\\
                \noalign{\smallskip}
                \hline
                \noalign{\smallskip}
 	\end{tabular}
\end{table}

\subsection{Fractions of galaxies in projected phase space}
In order to obtain correlated information on galaxy velocity and position, we study the projected phase space (PPS) diagram, i.e. $|V_{rf}|/\sigma$ versus $R/R_{200}$, as environment proxy and we explore the PPS distributions of the different galaxy populations at the group mass regime. Previous works have investigated the PPS for clusters alone \citep{Mahajan2011,Muzzin2014,Jaffe2015}, and for samples with both cluster and group mass halos \citep{Oman2013,Oman2016}. Our GAMA sample allows us to probe the group halo mass range alone ($10^{12}\leq (M_{200}/M_{\odot})< 10^{14}$) with a larger number of galaxies and to determine whether the segregation of star-forming and passive galaxies observed in the PPS of clusters also exists in groups.

\begin{figure*}
\includegraphics[width=0.50\textwidth]{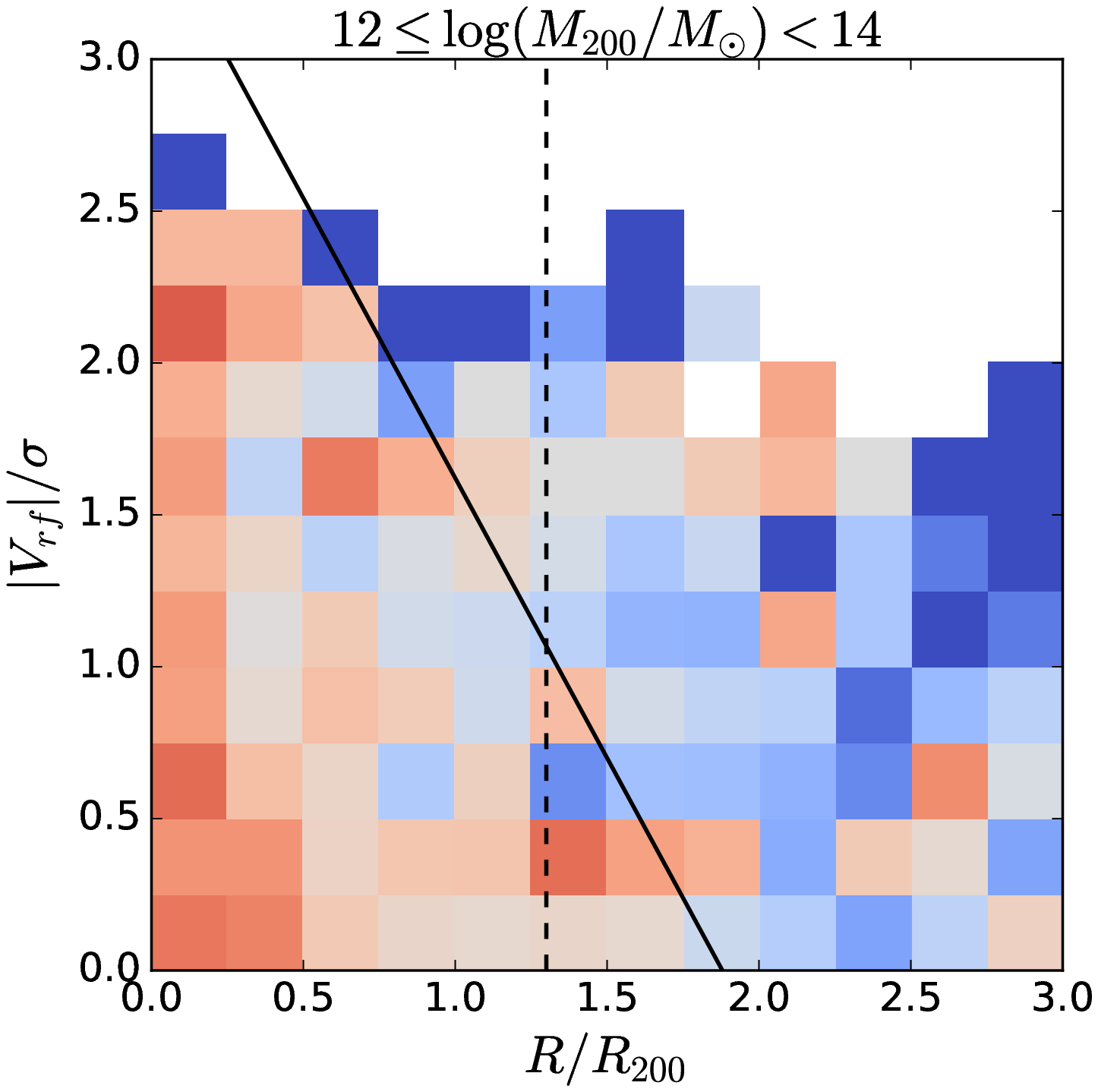}  
\includegraphics[width=0.50\textwidth]{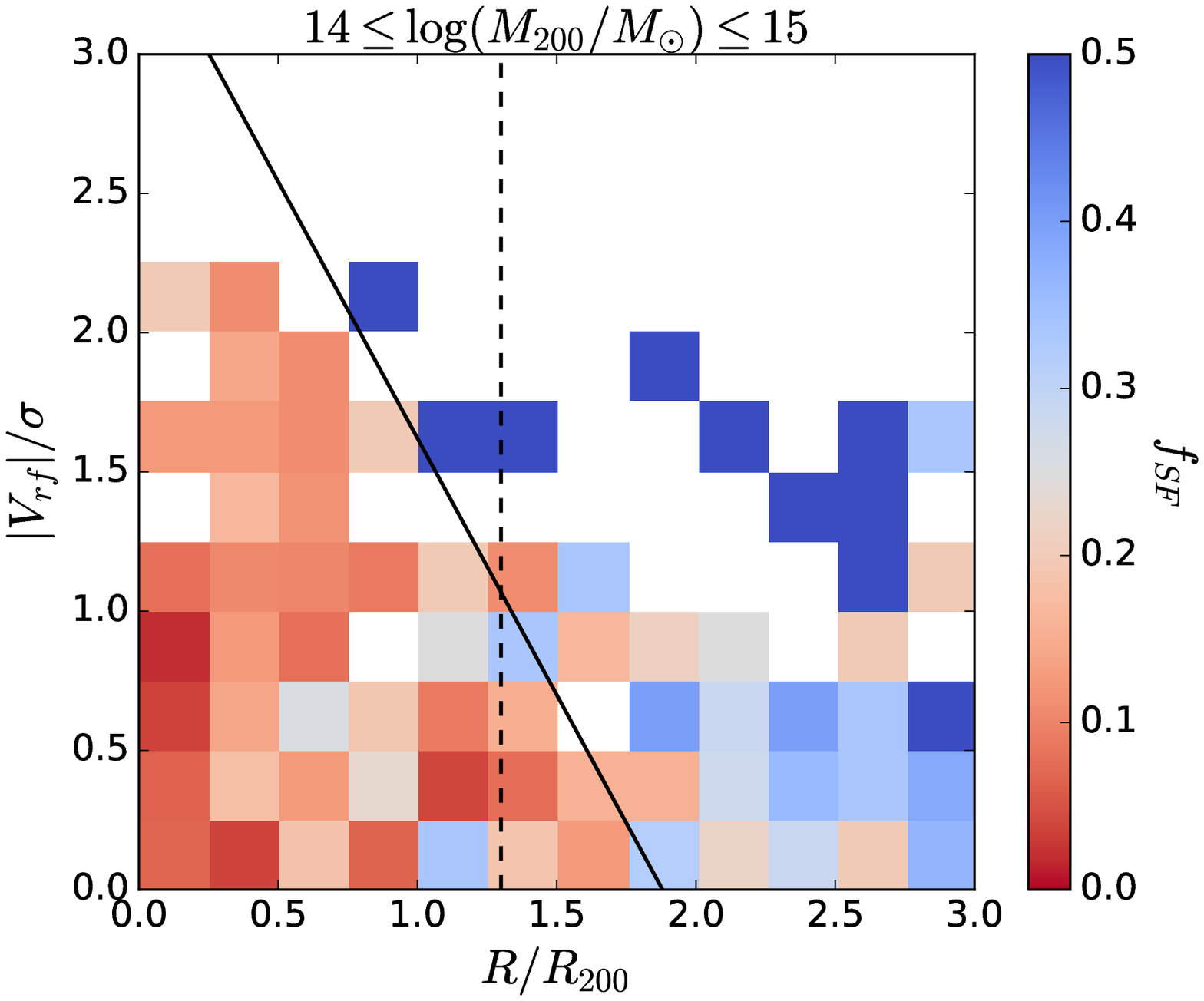}
  \caption{2D histograms of star-forming galaxy fractions binned in the PPS for groups and clusters. The virialized region at small radii is characterized by low values of $f_{\rm SF}$ (redder colors) and it is most populated by passive galaxies, whereas for the infalling region at large distances the fraction of star-forming galaxies is higher (bluer colors). The solid and dashed black lines represent the separation between the virialized and infalling galaxy populations found by \citet{Oman2013} and \citet{Mahajan2011}, respectively.}
\label{2Dhistogram}  
\end{figure*}

We use the Restricted Sample and Figure~\ref{2Dhistogram} shows the 2D histograms of star-forming fractions binned in the PPS. We consider the radial range out to 3$\,R_{200}$ in order to study a region containing galaxies which can be or have been physically affected by the group/cluster environment. We plot the separation line (solid black) between the region that are likely to have a high fraction of recently accreted galaxies (i.e., the infalling population) and the region with galaxies that have been inside the group/cluster for an extended period of time (i.e., the virialized population), found by \citet{Oman2013} for simulated groups and clusters. We adapt the galaxy velocity$-$radius relation of \citet{Oman2013} to:
\begin{equation}
\frac{|V_{\rm rf}|}{\sigma}=-\frac{4}{3}\frac{1}{1.25} \frac{R}{R_{200}}\sqrt{3}+2\sqrt{3}  
\end{equation}
where the factors of $\sqrt{3}$ and 1.25 convert from the 3D velocity dispersion $\sigma_{\rm 3D}$ and the virial radius $R_{\rm vir}$ used by \citet{Oman2013} to our 1D $\sigma$ and $R_{200}$, i.e. $\sigma=\sigma_{\rm 3D}/\sqrt{3}$ and $R_{\rm vir}=1.25\,R_{200}$ \footnote{In the relation $R_{\rm vir}=1.25\,R_{200}$ the value 1.25 does not compare in \citet{Oman2013}, but it has been provided to us via private communication from Mike Hudson.}
at $z=0$. We also plot the separation line (dashed black) between the virialized and infalling populations established by \citet{Mahajan2011} at $R\sim1.3\,R_{200}$.

The virialized region at small radii is characterized by low values of $f_{\rm SF}$ and it is most populated by passive galaxies, whereas for the infalling region at large distances the fraction of star-forming galaxies is higher. This result is observed for both groups and clusters. We conclude that the segregation of star-forming and passive galaxies in the PPS already detected in clusters and in combined group and cluster samples is also observable in low-mass halos alone.

\subsection{Segregation in velocity space}
The analysis of the star-forming and passive fractions in the radial space reveals a segregation of the two galaxy populations in both groups and clusters. However, previous works focusing on clusters also observed galaxy color/spectral type and luminosity segregation when considering velocity space alone (e.g., \citealp{Biviano1992,Biviano1997,Adami1998,Ribeiro2013,Haines2015,Barsanti2016}). These effects have been also detected in groups, but are less studied (e.g., \citealp{Girardi2003,Lares2004,Ribeiro2010}).

\begin{figure*}
\centering
\includegraphics[scale=0.37]{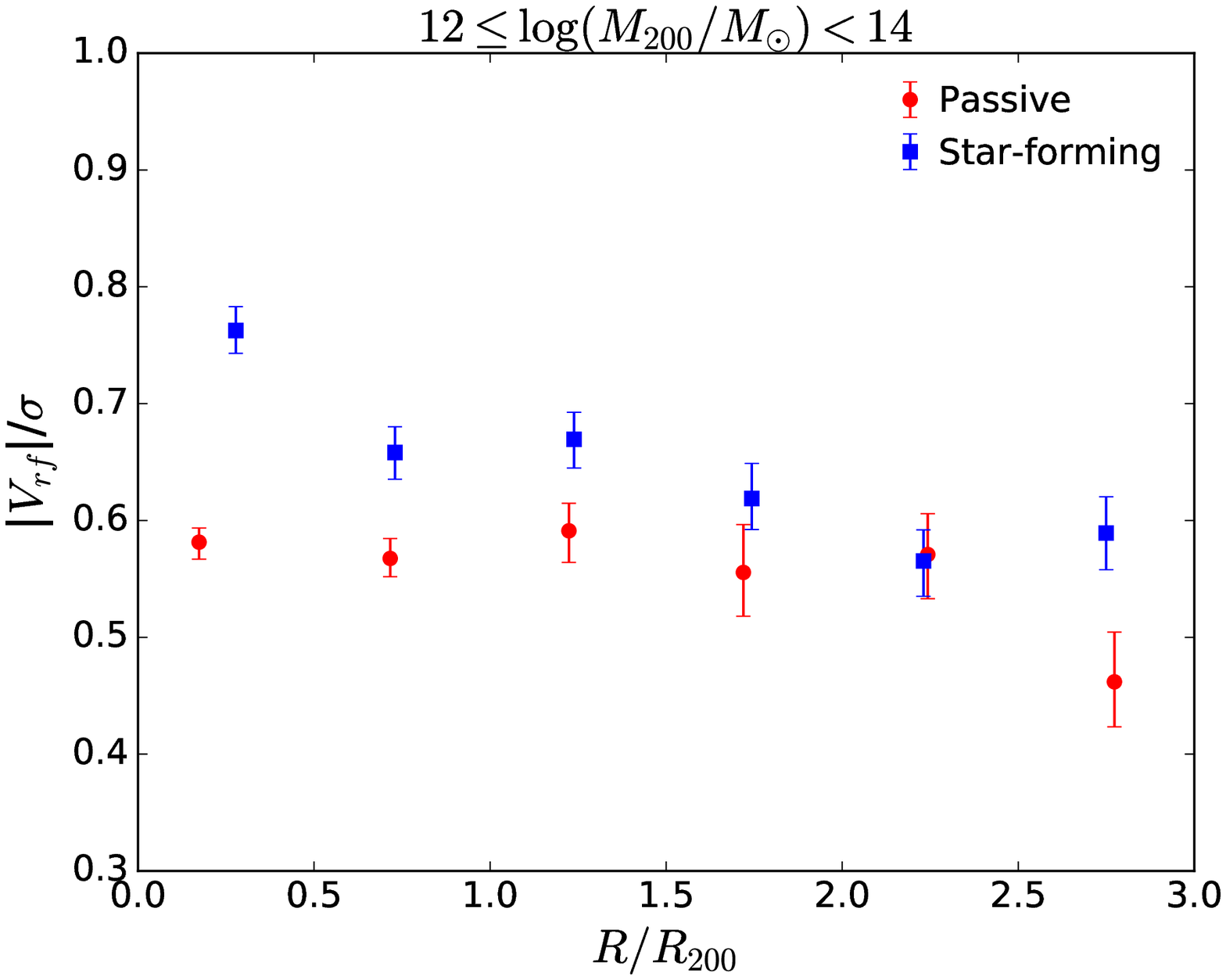}
\includegraphics[scale=0.37]{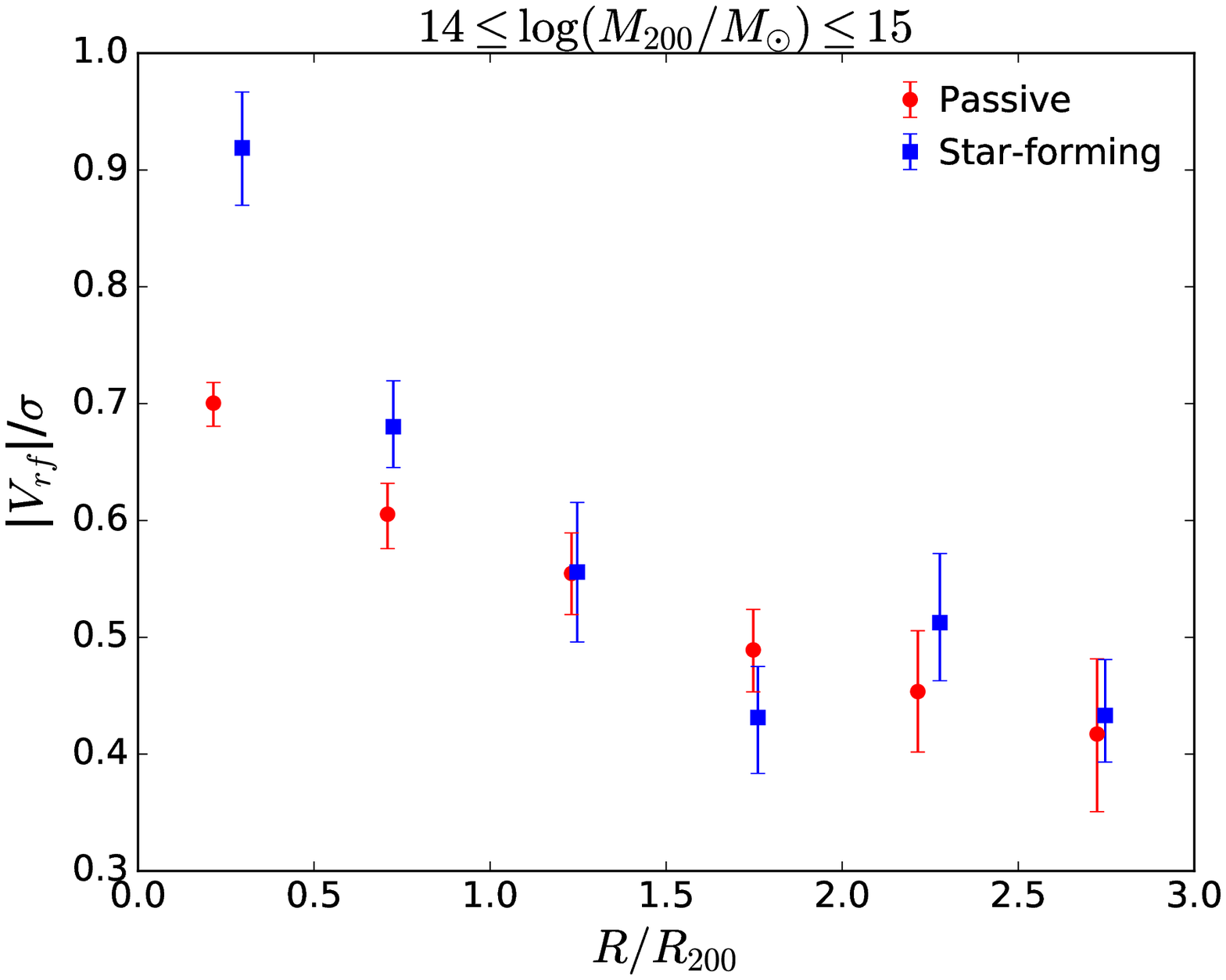}
\caption{$|V_{\rm rf}|$/$\sigma$ versus $R/R_{200}$ for star-forming and passive galaxies within $3\,R_{200}$ in groups and clusters. We plot median values in 6 radial bins and bootstrap errors at 68\% c.l.; the abscissa points are set to the biweight mean of the $R/R_{200}$ distribution within the bin of interest. Star-forming galaxies within $1.5\,R_{200}$ have higher $|V_{\rm rf}|$/$\sigma$ values when compared with the passive galaxy population.}
	\label{VelSegreg}
\end{figure*}

We consider the Full Sample and analyze the kinematics of galaxies, comparing the velocity profiles of the different populations as a function of radius and galaxy stellar mass. Figure~\ref{VelSegreg} shows the median $|V_{\rm rf}|$/$\sigma$ versus $R/R_{200}$ plot within $3\,R_{200}$ to focus on the physically bound group/cluster region. There is a galaxy spectral type segregation in the velocity space: star-forming galaxies within $1.5\,R_{200}$ tend to have higher $|V_{\rm rf}|$/$\sigma$ values when compared with the passive galaxy population. 
In order to check if this difference is statistically significant we apply the $\chi^{2}-$test. For groups (clusters) we find that the $|V_{\rm rf}|$/$\sigma$ distributions of 2690 (620) star-forming and 5203 (1882) passive galaxies within $1.5\,R_{200}$ are different at the $\geq 99.99$\% confidence level (c.l.). We confirm the segregation of the passive and star-forming populations in the velocity space at both the group and cluster mass regimes.  

\begin{figure*}
	\centering
        \includegraphics[scale=0.37]{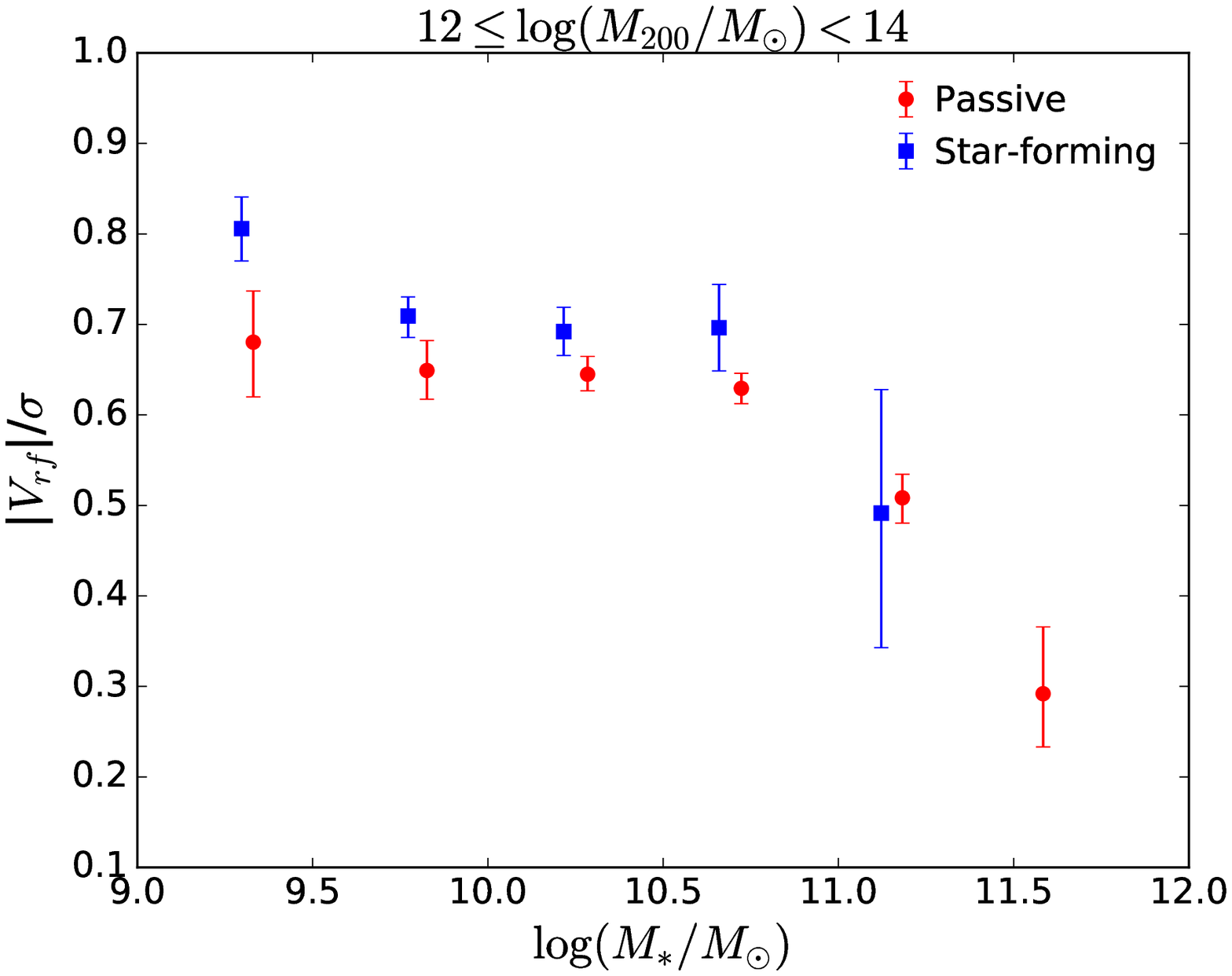}
        \includegraphics[scale=0.37]{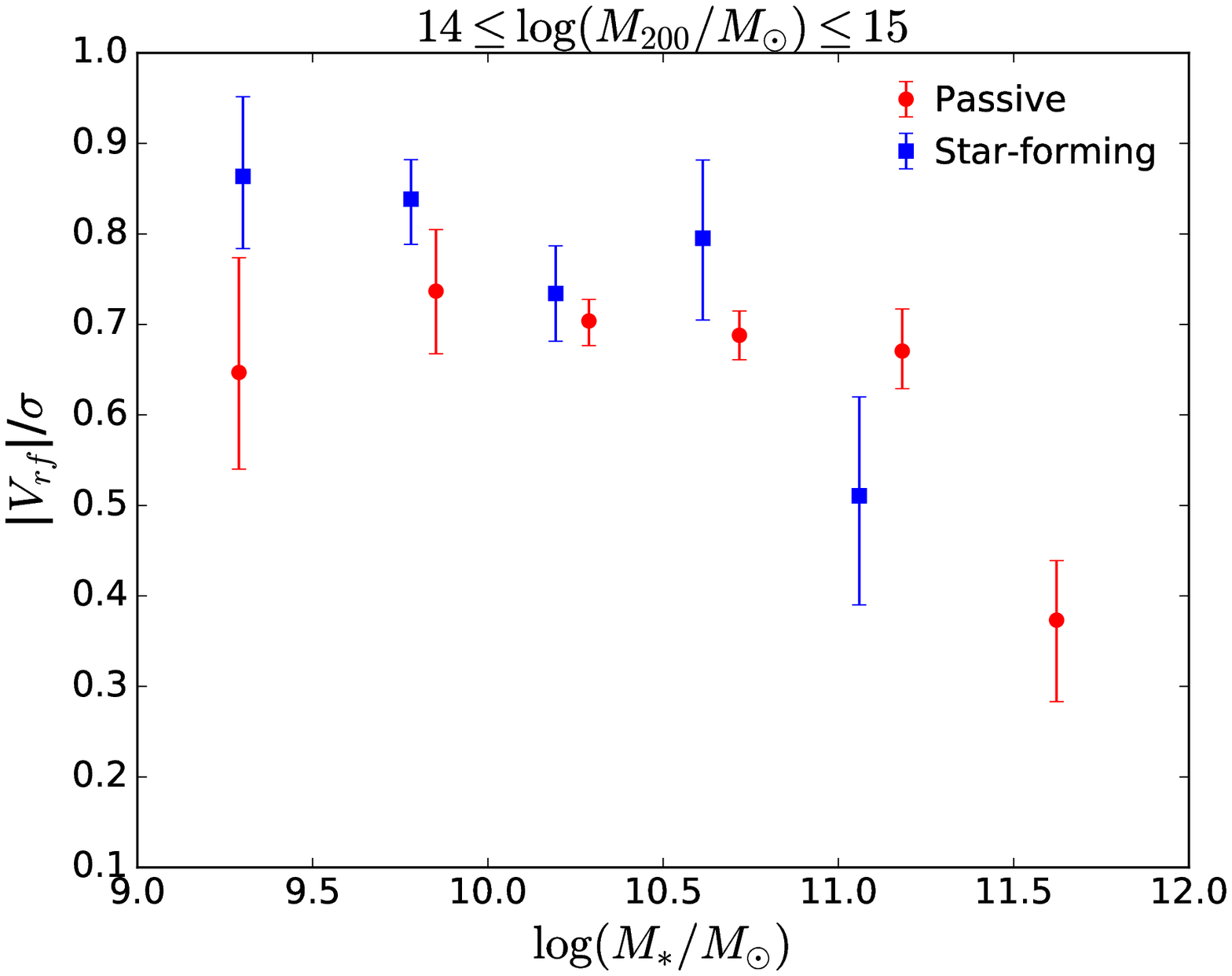}
	\caption{$|V_\text{rf}|$/$\sigma$ versus $M_{*}$ for star-forming and passive galaxies within $1\,R_{200}$ in groups and clusters. We plot median values in 6 bins and bootstrap errors at 68\% c.l.; the abscissa points are set to the biweight mean of the $M_{*}$ distribution within the bin of interest. More massive galaxies have lower $|V_{\rm rf}|$/$\sigma$ values than the low-mass ones which show a constant trend in velocity.}
	\label{MassSegreg}
\end{figure*}

Finally, we explore a possible galaxy stellar mass segregation in velocity space, since \citet{Kafle2016} observed no $M_{*}$ segregation with radius for GAMA group galaxies. We use the Full Sample and plot in Figure~\ref{MassSegreg} the median $|V_{\rm rf}|$/$\sigma$ versus $M_{*}$ for star-forming and passive galaxies of groups and clusters. We restrict this analysis within 1$\,R_{200}$ since this segregation is likely associated to secondary relaxation processes within the halo virialized regions \citep{Binney2008}. From the GAMA halos of the Full Sample we exclude the central galaxies which have been defined by \citet{Robo2011}. The inclusion of these galaxies could potentially bias our results. The selection of \citet{Robo2011} is generally robust, but it is based on an iterative procedure and it is possible that this method chooses the incorrect central galaxy, thus there may be some contamination. 

Massive passive galaxies show evidence of segregation in the velocity space: more massive galaxies have lower $|V_{\rm rf}|$/$\sigma$ values than the low-mass ones. However, this effect does not appear to be a statistically significant result for massive star-forming galaxies likely due to the low numbers of these galaxies with high $M_{*}$ \citep{Taylor2015}. For both the galaxy populations in groups, the trend in velocity remains approximately constant for galaxies with $10^{9}\leq (M_{*}/M_{\odot})< 10^{10.7}$ and then it decreases for $10^{10.7}\leq (M_{*}/M_{\odot})\leq 10^{12}$. For clusters the velocity decline starts at about $M_{*}\geq 10^{11.2} M_{\odot}$ and $M_{*}\geq 10^{10.7} M_{\odot}$ for passive and star-forming galaxies, respectively. To statistically evaluate this segregation, we apply the Spearman test in order to estimate the correlation between $|V_{\rm rf}|$/$\sigma$ and $M_{*}$ for galaxies with a flat trend and for those with a decline in velocity separately. Tables~\ref{MassSegregSpearman} reports the $P$-values for star-forming and passive galaxies in each $M_{200}$ range. Massive passive galaxies have smaller $P$-values implying a strong segregation. On the other hand, only the massive star-forming galaxies in clusters present a marginally statistically significant decrease in velocity. Finally, for both the galaxy populations low-mass galaxies do not show a correlation between $|V_{\rm rf}|$/$\sigma$ and $M_{*}$ and have a flat trend in velocity.

These results are in agreement with a scenario where the dynamical friction mechanism is able to slow the orbital motion of galaxies in groups and clusters \citep{Biviano1992,Adami1998,Girardi2003,Ribeiro2010,Ribeiro2013,Barsanti2016}. In agreement with the previous works, we confirm that this deceleration is a function of galaxy mass: for more massive galaxies, the higher the deceleration. We also observe that this segregation is stronger for passive galaxies than for star-forming galaxies.

\begin{table}
  \caption{Spearman test.}
  \centering
\label{MassSegregSpearman}
\begin{tabular}{l c r r r}
  \hline
  \noalign{\smallskip}
\hline
\noalign{\smallskip}
\multicolumn{1}{c}{\small{$(M_{200}/M_{\odot})$}}
&\multicolumn{1}{l}{\small{Type}}
&\multicolumn{1}{r}{\small{$\log(M_{*}/M_{\odot})$}}
&\multicolumn{1}{c}{\small{$N_{\rm gals}$}}
&\multicolumn{1}{c}{\small{$P$}}
\\
\hline
\noalign{\smallskip}
$10^{12}-10^{14}$ &\small{PAS}  &$9.0-10.7$ &2596 & 0.6533\\
$10^{12}-10^{14}$ &\small{PAS}  &$10.7-12.0$ &1942 &0.0004 \\
$10^{12}-10^{14}$ &\small{SF}  &$9.0-10.7$ &1942 &0.1260\\
$10^{12}-10^{14}$ &\small{SF}  &$10.7-12.0$ &111 &0.6689 \\
\noalign{\smallskip}
	\hline
        \noalign{\smallskip}
$10^{14}-10^{15}$&\small{PAS} &$9.0-11.2$&1031 &0.8876\\
$10^{14}-10^{15}$&\small{PAS} &$11.2-12.0$&604 &0.0313\\
$10^{14}-10^{15}$&\small{SF}  &$9.0-10.7$ &462 & 0.1312\\
$10^{14}-10^{15}$&\small{SF}  &$10.7-12.0$ &12 &0.0513\\
\noalign{\smallskip}
\hline
\noalign{\smallskip}
\end{tabular}
\tablecomments{$P$-values quantifying the correlation between $|V_{\rm rf}|$/$\sigma$ and $M_{*}$ for passive and star-forming galaxies in groups/clusters.}
\end{table}

\subsection{SFR$-$radius relation for star-forming galaxies}
The studies of the radial, projected phase and velocity spaces suggest that the star-forming group galaxies are recently accreted and represent an infalling population from the field to the halo. This result is well established for star-forming cluster galaxies which also show suppressed star formation with respect to the field (e.g., \citealp{Lewis2002,Gomez2003,vonderLinden2010,Paccagnella2016}). However, this latter observation is less clear for star-forming galaxies in groups. In the Sections 3.4$-$3.8 we explore whether and how the group environment affects the star formation properties, analyzing the dependence of SFR on radius and stellar mass.

We focus on star-forming galaxies of the Full Sample within 9$\,R_{200}$, probing a similar stellar mass and radial range as \citet{Ras2012} and including a benchmark sample of field galaxies (see Section 2.3). The star-forming galaxies are spectroscopically selected as described in the Section 2.4. For these galaxies we define in the Section 2.5 two different SFR estimators, i.e. $\rm SFR_{H\alpha}$ and $\rm SFR_{MAGPHYS}$, based on the spectroscopic and photometric properties of galaxies, respectively. Table~\ref{samplegalsSF} lists the number of star-forming galaxies with available $\rm SFR_{H\alpha}$ and $\rm SFR_{MAGPHYS}$. There are fewer star-forming galaxies with measured $\rm SFR_{H\alpha}$ when compared with those with available $\rm SFR_{MAGPHYS}$: 460 galaxies have no measured $\rm SFR_{H\alpha}$ because their spectra are not flux calibrated and/or it is not possible to make obscuration corrections.

\begin{table}
  \caption{Full Sample: star-forming galaxies out to 9$\,R_{200}$.}
  \centering
  \label{samplegalsSF}
	\begin{tabular}{l r c}
	  \hline
          \noalign{\smallskip}
	  \hline
          \noalign{\smallskip}
		\multicolumn{1}{l}{$(M_{200}/M_{\odot})$}
                &\multicolumn{1}{c}{$N_{\rm SFR, H\alpha}$}
                &\multicolumn{1}{c}{$N_{\rm SFR, MAGPHYS}$}
		\\
		\hline
                \noalign{\smallskip}
                $10^{12}-10^{14}$ & 7213& 7568\\
		$10^{14}-10^{15}$ & 2565&2670 \\
                \noalign{\smallskip}
		\hline
                \noalign{\smallskip}
	\end{tabular}	
\end{table}

\begin{figure}
  \centering
\resizebox{\hsize}{!}{\includegraphics{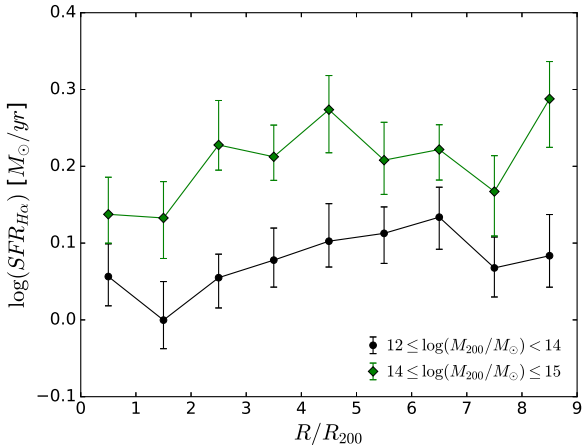}}
\caption{$\rm SFR_{H\alpha}$ as a function of projected group/cluster-centric distance for star-forming galaxies. We plot median values binned every 1$\,R_{200}$ with errors at the 68\% c.l. There is a decline of $\rm SFR_{H\alpha}$ towards the halo inner regions.}
\label{SFRvsRadius}
\end{figure}

Figure~\ref{SFRvsRadius} shows median $\rm SFR_{H\alpha}$ values versus $R/R_{200}$ for star-forming galaxies associated to groups and clusters. For clusters the $\rm SFR_{H\alpha}$ remains constant over $2.5< (R/R_{200}) \leq 9$ and then decreases towards the cluster center. For groups there is a continuous decrease of the star formation activity from the group outskirts to the inner regions. The shift towards higher $\rm SFR_{H\alpha}$ for clusters with respect to groups is due to the fact that at higher redshift we observe halos with higher mass and SFR (see the \textit{upper panel} of Figure~\ref{histogramsZM200}).

\begin{figure}
  \centering
\resizebox{\hsize}{!}{\includegraphics{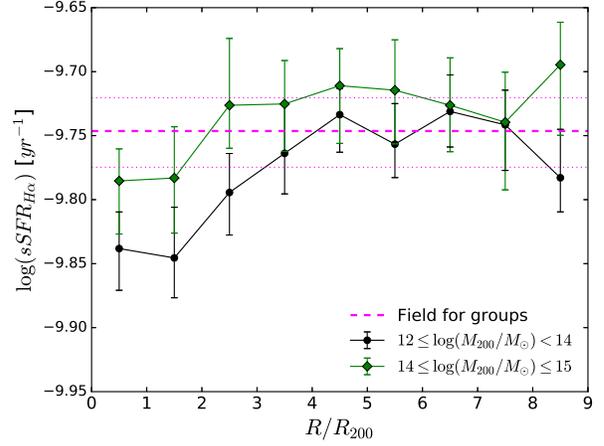}}
\caption{$\rm sSFR_{H\alpha}$ as a function of projected group/cluster-centric distance for star-forming galaxies. The dashed magenta line indicates the median $\rm sSFR_{H\alpha}$ for galaxies associated to groups but outside 4.5$\,R_{200}$ and defined as belonging to the field, with the dotted magenta lines marking the uncertainties on that. The $\rm sSFR_{H\alpha}$ declines by a factor $\sim1.2$ from the field to the group inner region.}
 	\label{sSFRvsRadius}
\end{figure}

We also analyze the dependence of specific star formation rate (sSFR=SFR/$M_{*}$) on radius. Figure~\ref{sSFRvsRadius} confirms the decline of $\rm SFR_{H\alpha}$ at $\sim2.5\,R_{200}$ towards the cluster centers and it shows a drop in $\rm sSFR_{H\alpha}$ at $\sim4.5\,R_{200}$ for star-forming galaxies in groups, which is not evident in Figure~\ref{SFRvsRadius} where there is a continuous decline. In order to compare the star formation in the group environment with that in the field, we consider star-forming galaxies at $R\leq 4.5\,R_{200}$ as group members and those with $R>4.5\,R_{200}$ as field galaxies. Comparing the median $\rm sSFR_{H\alpha}$ value for the field (magenta line) with that for star-forming group galaxies in the $0 \leq (R/R_{200})< 1$ bin (black point), it can be seen that the $\rm sSFR_{H\alpha}$ declines by a factor of $\sim1.2$. These results are in agreement with the outcome of \citet{Ras2012}, who found a decrease in the sSFR as a function of the projected group-centric distance for star-forming galaxies in nearby groups.

\begin{figure}
  \centering
\resizebox{\hsize}{!}{\includegraphics{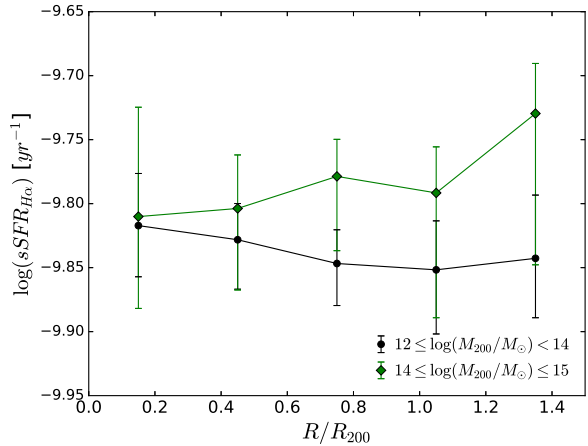}}
\caption{$\rm sSFR_{H\alpha}$ versus $R/R_{200}$ for star-forming galaxies out to $1.5\,R_{200}$ in groups and clusters. There is no change of $\rm sSFR_{H\alpha}$ with radius.}
\label{sSFRvsRadiusZip}
\end{figure}

In order to compare our analysis with that of \citet{Ziparo2013}, we consider galaxies only out to $1.5\,R_{200}$ in Figure~\ref{sSFRvsRadiusZip}, since \citet{Ziparo2013} investigated the SFR$-$ and sSFR$-$radius relation within this radial distance. Figure~\ref{sSFRvsRadiusZip} shows that there is no statistically significant correlation between $\rm sSFR_{H\alpha}$ and $R/R_{200}$ at small radii for groups, but the uncertainties are large. This results agrees with \citet{Ziparo2013}.


The decline in star formation towards the inner group/cluster-centric radii is present when we consider the group/cluster infalling regions and their surroundings out to $9\,R_{200}$, in agreement with \citet{Ras2012}. However, similar to \citet{Ziparo2013} there is no decline in sSFR for group/cluster galaxies within $1.5\,R_{200}$ which is the region most affected by the group/cluster environment. This might indicate that the SFR of group galaxies is stopped slowly, since the suppression is too slow to be detected at small radii. As a consequence, the quenching timescale probably is of the order of few Gyr and comparable to the group crossing time. This result is in agreement with the conclusion of \citet{vonderLinden2010}, who proposed a scenario in which star formation is quenched slowly. 

\begin{figure*}
\centering
\includegraphics[scale=0.37]{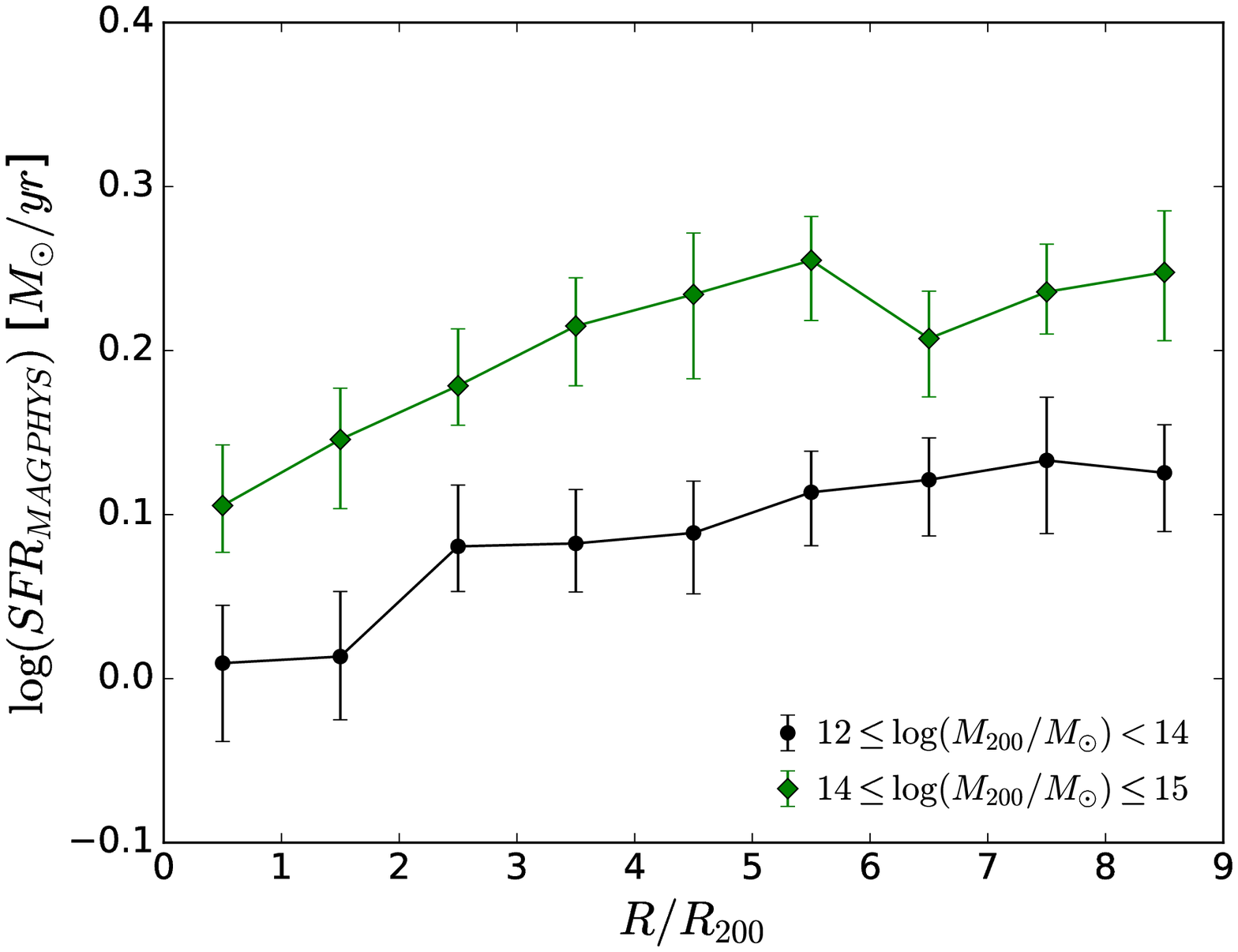}
\includegraphics[scale=0.37]{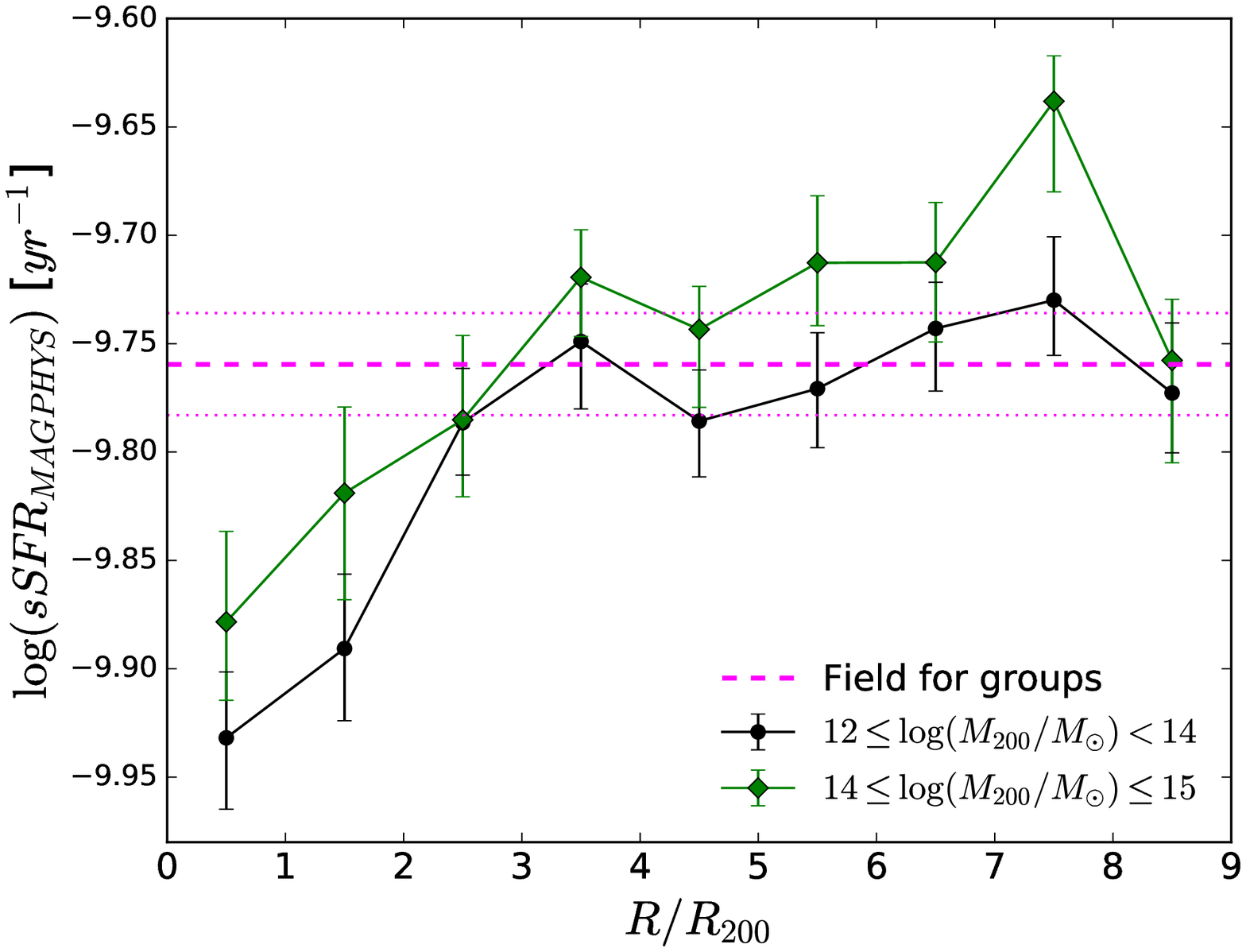}
\caption{$\rm SFR_{MAGPHYS}$ and $\rm sSFR_{MAGPHYS}$ versus $R/R_{200}$ for star-forming galaxies (\textit{left} and \textit{right panel}, respectively) in groups and clusters. We plot median values binned every 1$\,R_{200}$ with errors at the 68\% c.l. The dashed magenta line indicates the median $\rm sSFR_{MAGPHYS}$ for galaxies associated to groups but outside 4.5$\,R_{200}$ and defined as belonging to the field, with the dotted magenta lines marking the uncertainties on that. There is a decline in star formation with decreasing radius and $\rm sSFR_{MAGPHYS}$ declines by a factor $\sim1.5$ from the field to the group inner region.}
\label{SFRvsRadiusPhoto}
\end{figure*}

Finally, we explore the dependence of $\rm SFR_{MAGPHYS}$ and $\rm sSFR_{MAGPHYS}$ on projected group-centric radius for star-forming galaxies in Figure~\ref{SFRvsRadiusPhoto}. The \textit{left panel} shows for clusters a continuous decrease in $\rm SFR_{MAGPHYS}$ towards the cluster center, whereas for groups there is a drop at $\sim2.5\,R_{200}$. The decline in star formation is more evident in the \textit{right panel} which illustrates a decreasing trend in $\rm sSFR_{MAGPHYS}$ at $\sim 3.5\,R_{200}$ for both groups and clusters. Comparing the median $\rm sSFR_{MAGPHYS}$ value for star-forming group galaxies in the $0 \leq (R/R_{200})< 1$ bin (black point) with that for the associated field galaxies at $4.5 < (R/R_{200})\leq 9$ (magenta line), it can be seen that the $\rm sSFR_{MAGPHYS}$ declines by a factor of $\sim1.5$ from the field to the group inner region. The results obtained using the SED-fitting code MAGPHYS as a star formation estimator are in agreement with those based on the H$\alpha$ emission lines within the uncertainties.

Throughout the different analyses of this Section we have used three different radial limits, i.e. $1.5\,R_{200}$, $4.5\,R_{200}$, and $9\,R_{200}$. The primary motivation behind selecting these limits was to allow us to compare with previous studies such as \citet{Ras2012} and \citet{Ziparo2013}. However, these radial limits may also be interpreted in a more physical manner. The region $R\leq 1.5\,R_{200}$ is where a large fraction of galaxies that have encountered the group/cluster core reside \citep{Gill2005,Mahajan2011}, and where we observe the strongest signature of suppressed star formation. The range $1.5 \leq (R/R_{200})< 4.5$ is the region containing bound populations that may infall onto the group/cluster, in agreement with \citet{Rines2013} who found that the maximum radius enclosing gravitationally bound galaxies to the halo is at $4-5\,R_{200}$. This region is mainly populated by star-forming galaxies. Finally, beyond $4.5\,R_{200}$ there is the unbound field population.

In conclusion, a decline is observed in star formation activity with decreasing group-centric radius for star-forming galaxies. The radius at which this decline begins differs for the various measures of SFR and for the different halo mass ranges probed. Generally, the decline begins in the radial range $2.5\leq (R/R_{200}) \leq 4.5$. This distance is well beyond the radius at which the group environment is expected to play a role in quenching star formation, and is also beyond the apocentric distance to which a galaxy will travel after its first passage of the group \citep{Gill2005}. Thus, this may indicate that galaxies are being pre-processed in very low-mass groups not detected in the GAMA catalog, or in the filament environment (e.g., \citealp{Alpaslan2016}), prior to falling into the GAMA-identified halos. However, many factors can conspire to spread the point at which the SFR decline starts out in radius. For example, uncertainties in the estimate of $R_{200}$ are likely to be relatively high due to the propagation of the errors in the velocity dispersion/mass measurements used to define the $R_{200}$ values. These uncertainties will broaden any sharp decline in radius.

\subsection{sSFR histograms}
The $\rm sSFR_{H\alpha}-$radius relationship in Figure~\ref{sSFRvsRadius} shows that for groups the median $\rm sSFR_{H\alpha}$ declines with radius towards the group center, but is relatively flat for $R>4.5\,R_{200}$. This behavior at large radius is to be expected, since galaxies at these radii are too distant to have been affected by the group environment, and are highly unlikely to have traversed the group. Thus, we use the star-forming galaxies with $R>4.5\,R_{200}$ as a benchmark field sample for comparison to group member galaxies at $0\leq (R/R_{200})\leq 4.5$.

A consequence of the sSFR suppression in halos is that there may be a low sSFR galaxy population in cluster and group environments that may not be seen in the field, or there may be an overall shift in the SFR of all star-forming galaxies. In this context, we investigate the sSFR distributions in clusters, groups and in the field in order to check for possible differences. We investigate $\rm sSFR_{H\alpha}$ in order to avoid the effect of the known stellar mass relationship with SFR and since we are interested in comparing measurements of star formation from the last $\sim$ 10 Myrs. We analyze sSFR histograms since the median distills a lot of information about the distributions of the sSFRs at a given radius into one point. The median of the sSFR distribution of group galaxies can be different with respect to that of field galaxies because there can be an overall shift in the total distribution of groups with respect to the field. However, another reason is that the sSFR distribution of groups can have a significant asymmetry, or bi-modality, that produces a difference in the medians. Investigating the sSFR histograms, we explore if there is a difference of the sSFR distribution in groups with respect to that in the field and we try to understand which of the above reasons for different medians is the case. Discerning between these two causes is important because it may give clues about the mechanisms responsible for the sSFR quenching.

We consider the Full Sample. Group/cluster members are defined as galaxies with $0\leq (R/R_{200})\leq 4.5$ and having $V_{\rm rf}$ lower or equal to the infall velocities (see Figure~\ref{PPSmassrange}). We build the field as populated by galaxies with $4.5< (R/R_{200})\leq 9$ in and outside the curves representing the infall velocities, i.e. with $-5\leq (V_{\rm rf}/\sigma)\leq 5$, in order to obtain a statistically high number of field galaxies. Moreover, each field galaxy is assigned to a halo according to the procedure of \citet{Smith2004} described in Section 2.3. We produce two separate field samples for each of the $M_{200}$ range. Finally, the star-forming galaxy population in each field is selected by applying the same method described in Section 2.5 and used to define star-forming members. The numbers of star-forming group and cluster galaxies in each radial bin and in the respective field are reported in Table~\ref{tableSFRnumbers}. 

\begin{figure*}
  \centering
  \includegraphics[scale=0.36,angle=90]{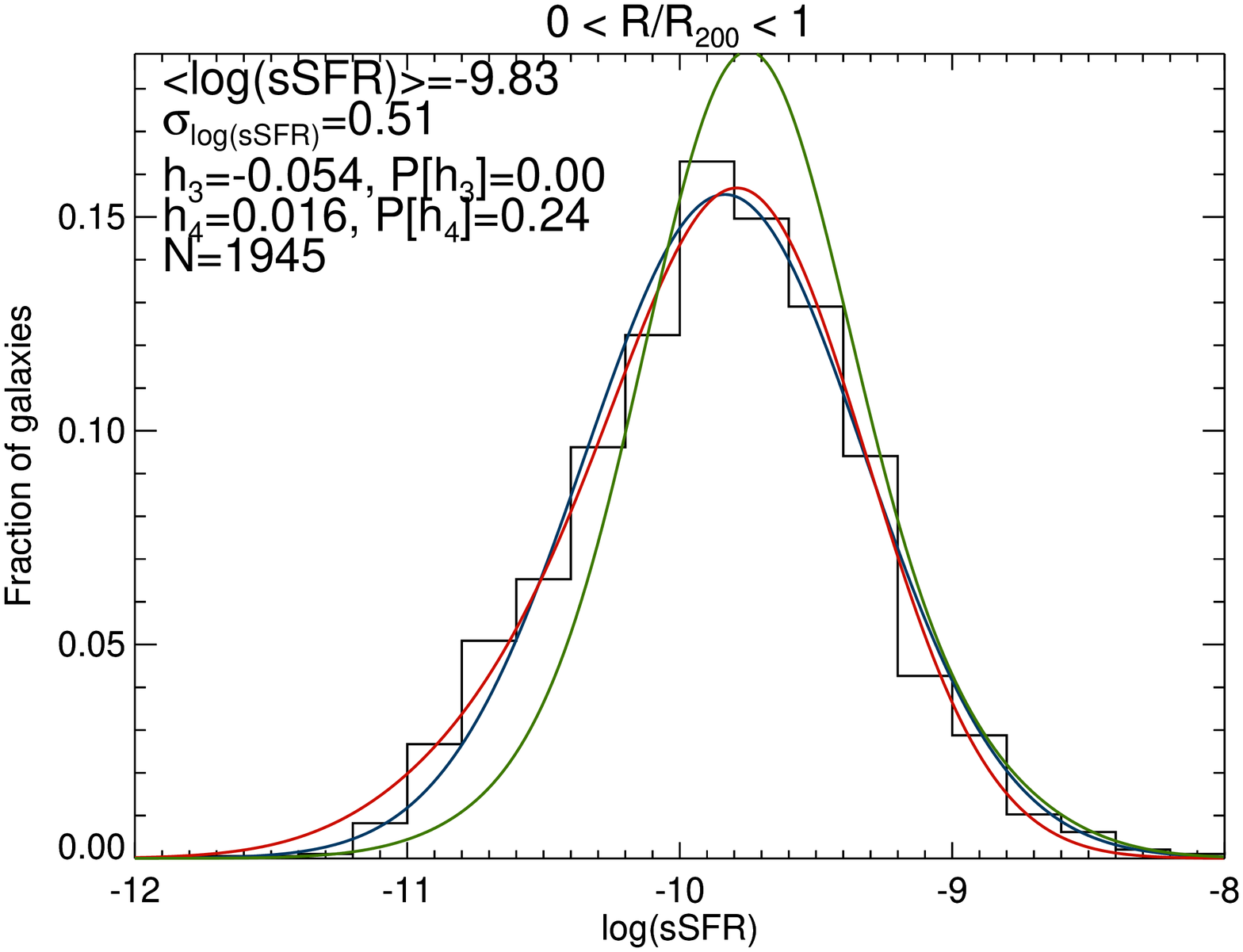}
  \includegraphics[scale=0.36,angle=90]{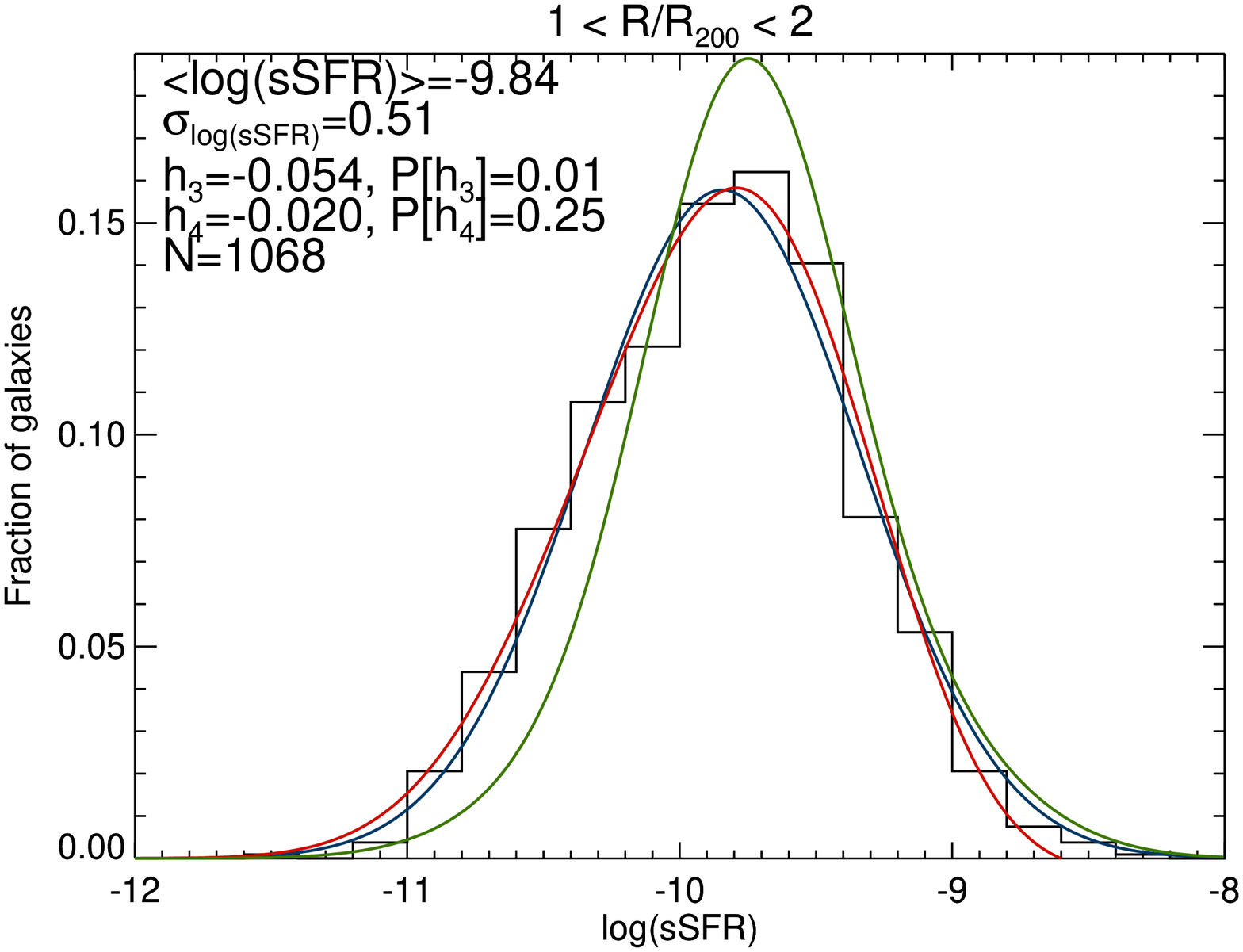}
 \includegraphics[scale=0.36,angle=90]{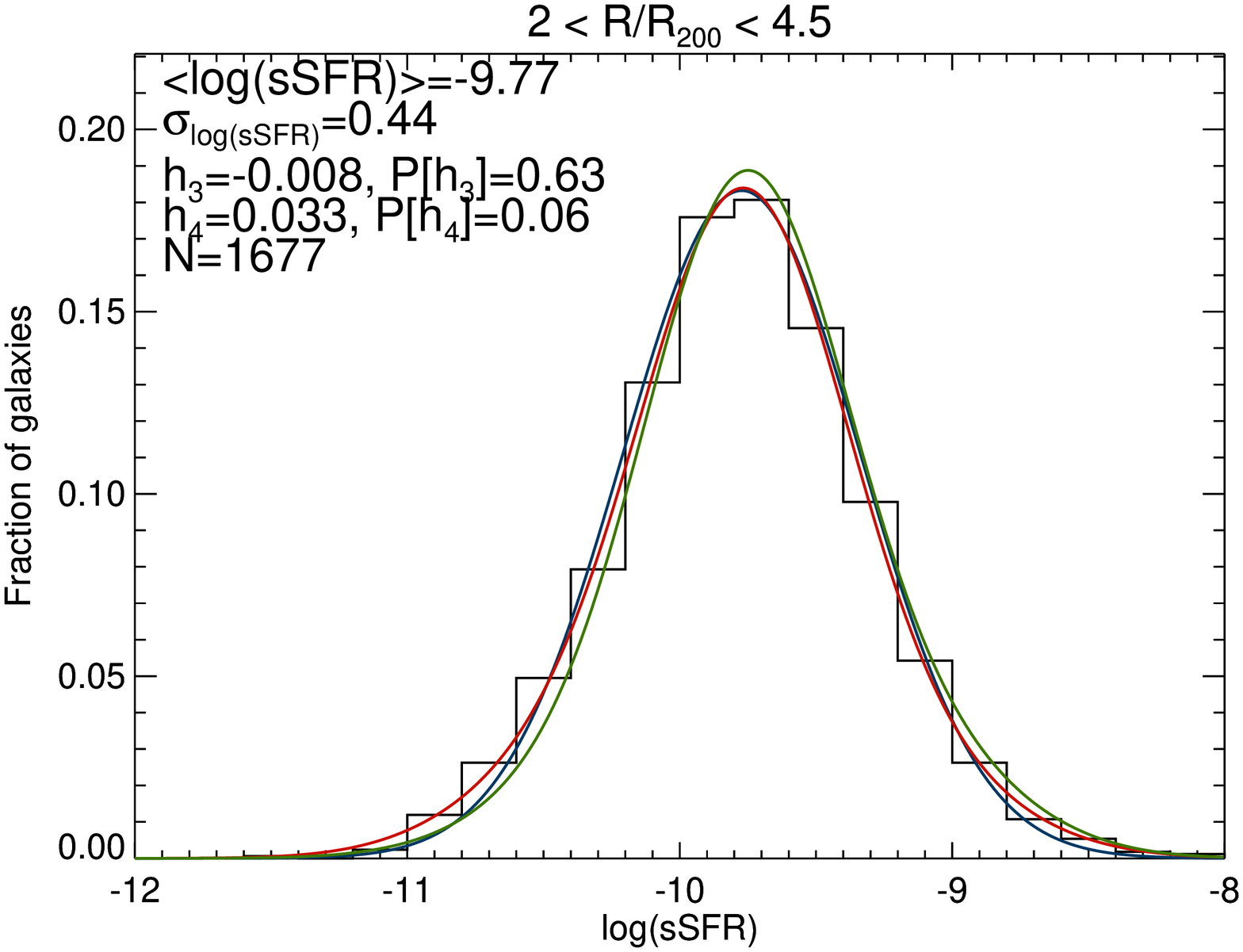}
 \includegraphics[scale=0.36,angle=90]{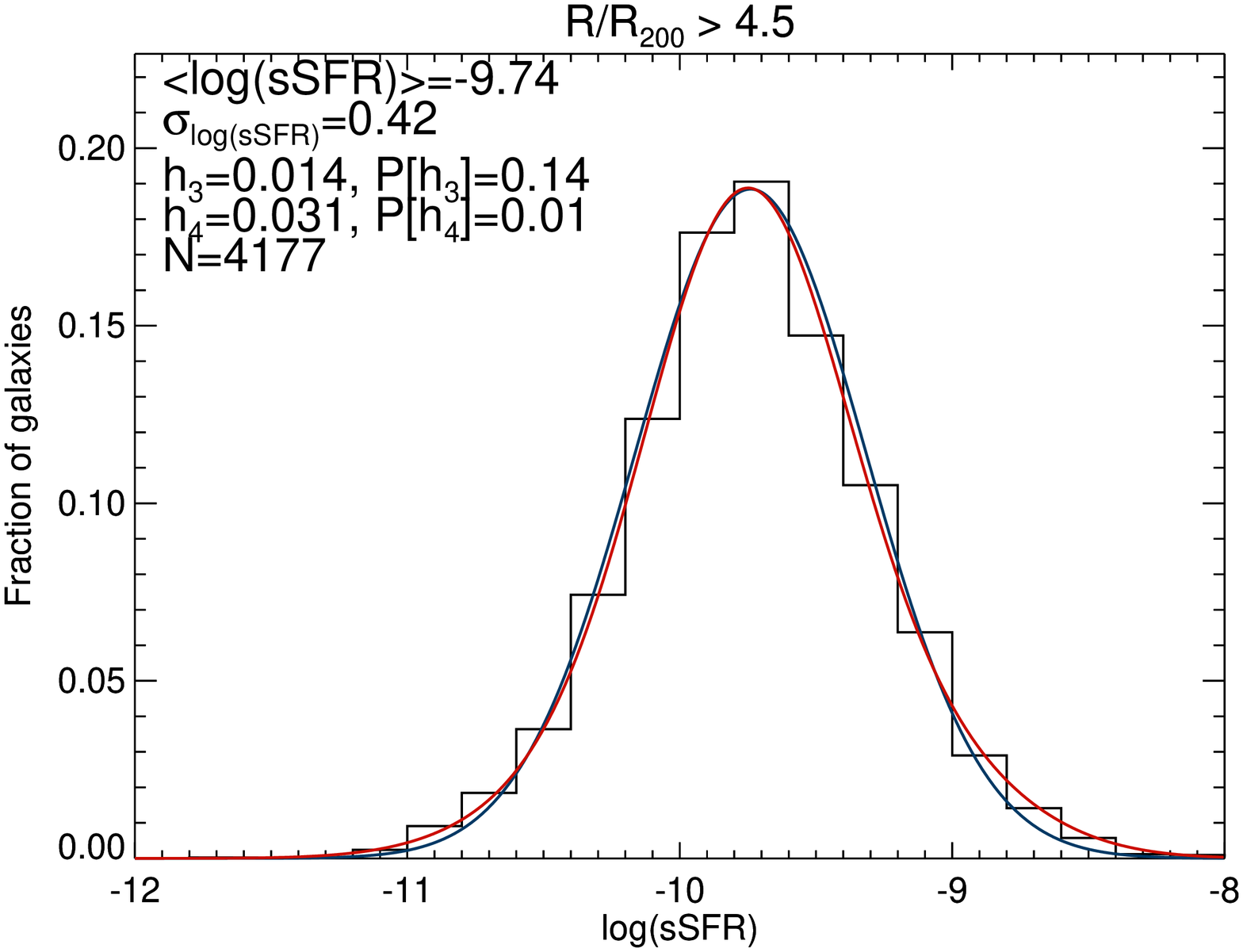}
\caption{ $\log(\rm sSFR_{H\alpha})$ histograms for the group star-forming galaxies in different radial ranges and in the respective field (black line), approximated by a series of Gauss-Hermite functions up to order 4 (red line) to estimate the asymmetric and symmetric departures from a Gaussian shape (blue line). The group $\log(\rm sSFR_{H\alpha})$ distributions are compared with the respective field one (green line), showing larger $\sigma_{\log(\rm sSFR_{H\alpha})}$ and lower $<\log(\rm sSFR_{H\alpha})>$ values than the field. The group histograms in the radial ranges $0\leq (R/R_{200}) <1$ and $1\leq (R/R_{200}) <2$ show significant evidence for asymmetry towards the lower $\rm sSFR_{H\alpha}$ side of the distribution relative to a Gaussian shape.}
\label{SFRhistogramsgroups}
\end{figure*}

\begin{figure*}
  \centering
  \includegraphics[scale=0.36,angle=90]{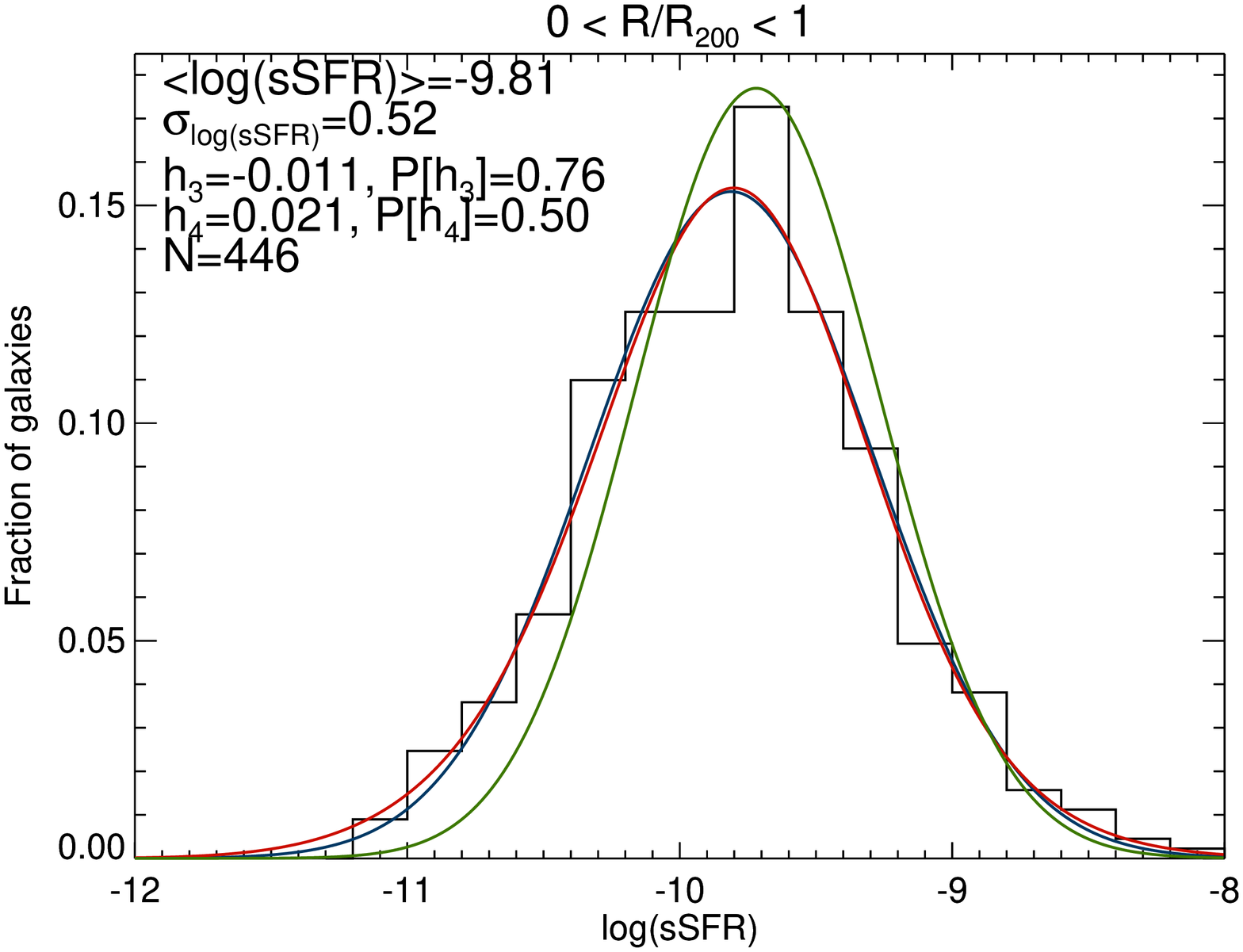}
  \includegraphics[scale=0.36,angle=90]{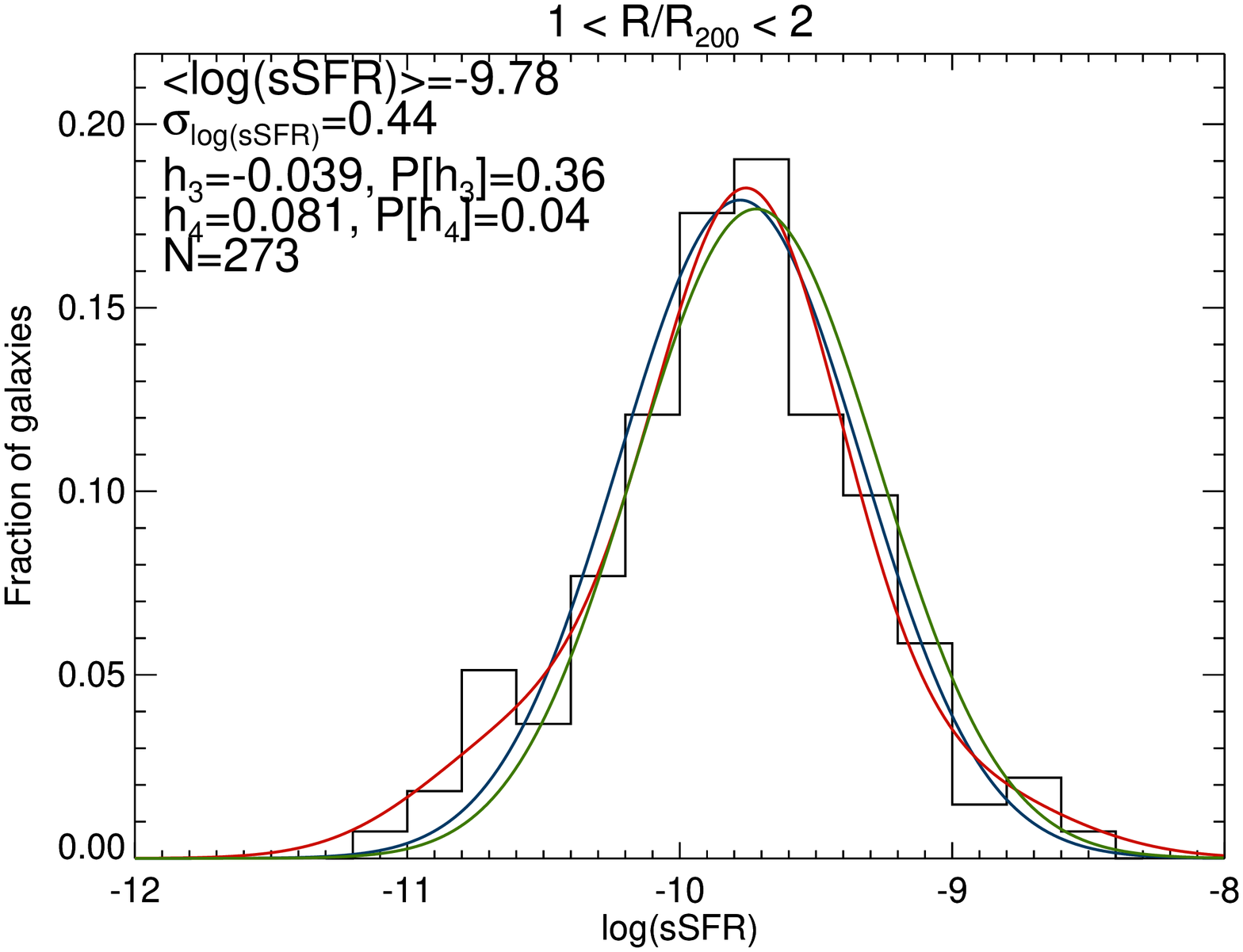}
 \includegraphics[scale=0.36,angle=90]{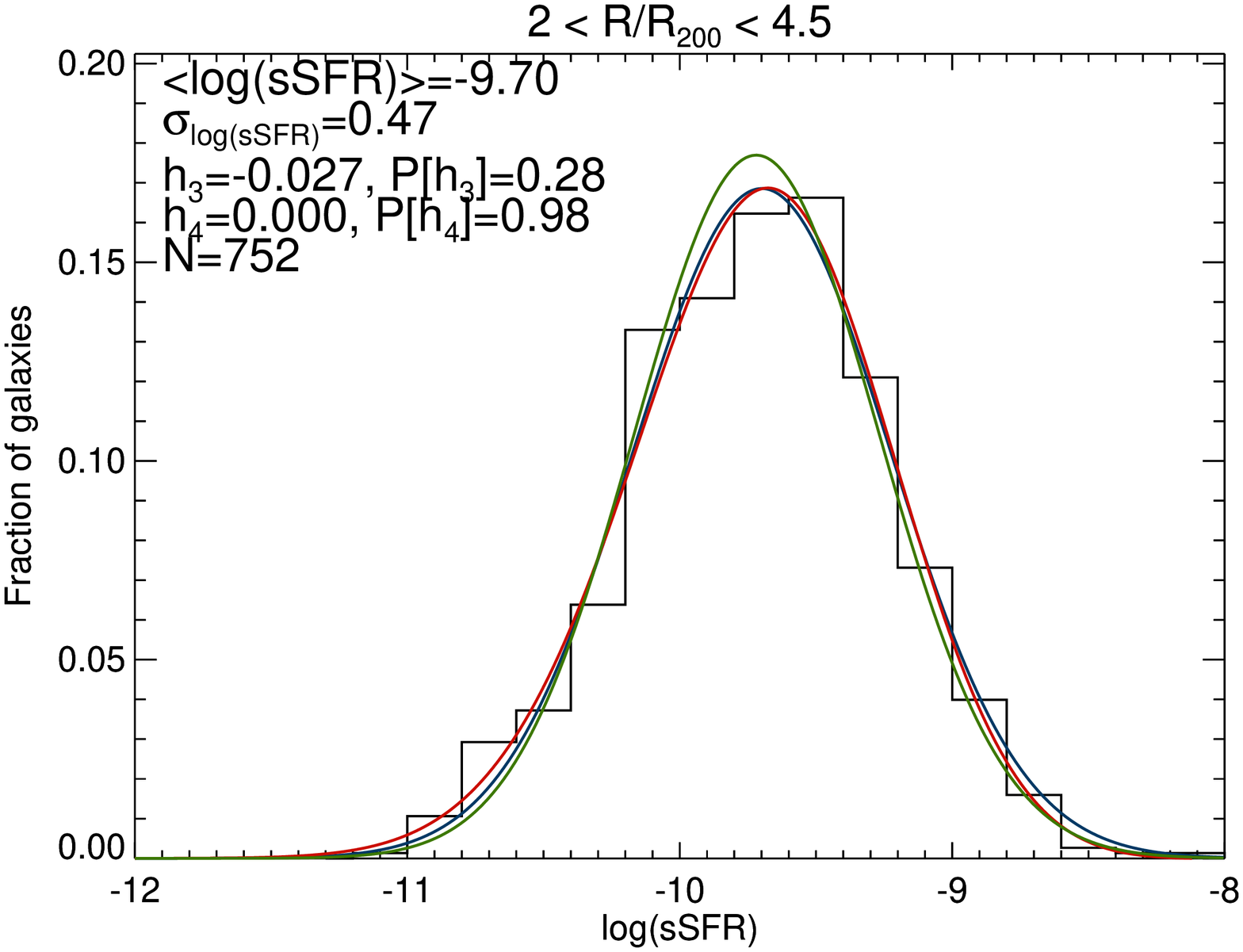}
 \includegraphics[scale=0.36,angle=90]{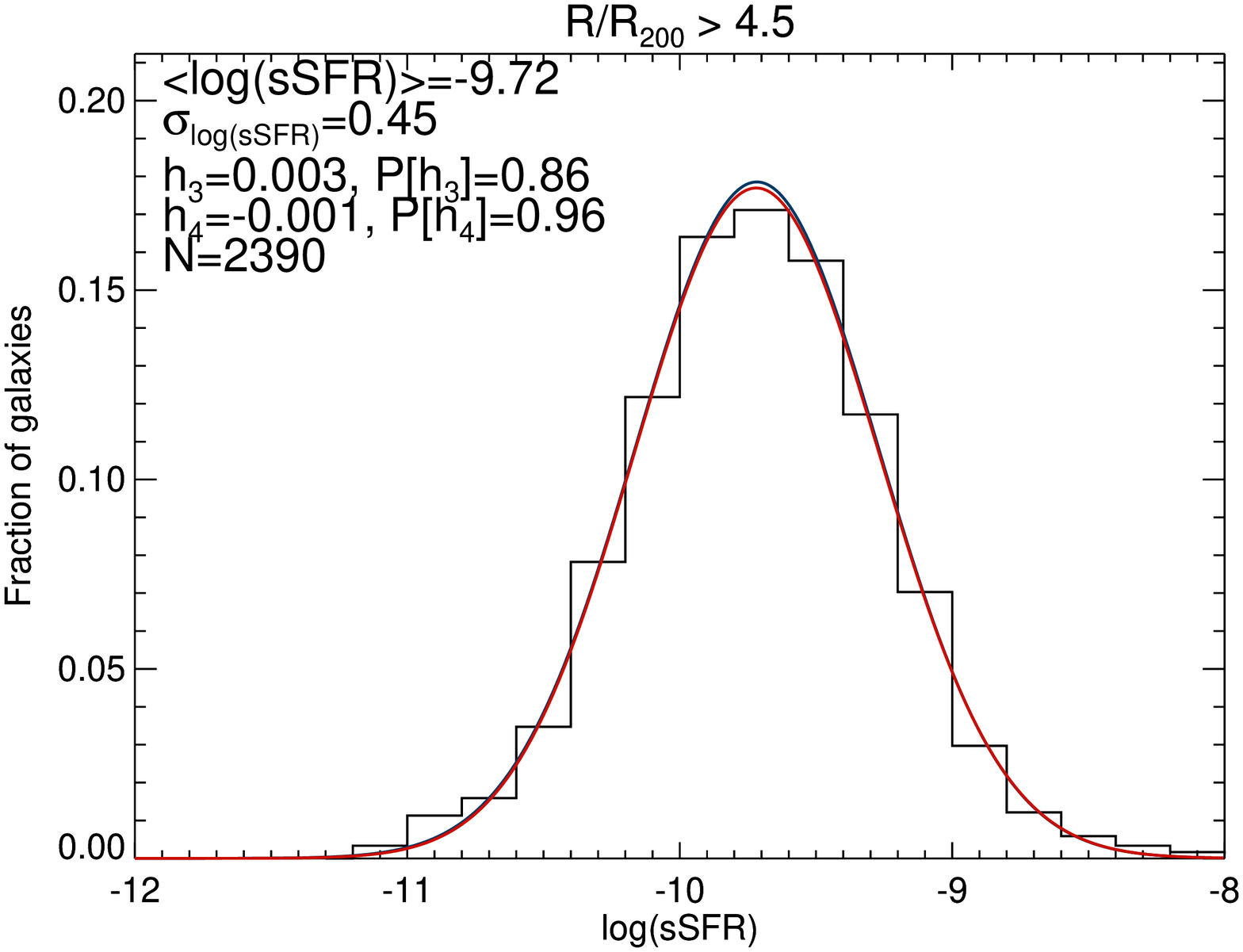}
\caption{ $\log(\rm sSFR_{H\alpha})$ histograms for the cluster star-forming galaxies in different radial ranges and in the respective field (black line), approximated by a series of Gauss-Hermite functions up to order 4 (red line) to estimate the asymmetric and symmetric departures from a Gaussian shape (blue line). The cluster $\log(\rm sSFR_{H\alpha})$ distributions are compared with the respective field one (green line), showing larger $\sigma_{\log(\rm sSFR_{H\alpha})}$ and lower $<\log(\rm sSFR_{H\alpha})>$ values than the field.}
\label{SFRhistogramsclusters}
\end{figure*}

We apply the Kolmogorov$-$Smirnov test (K$-$S test; \citealt{Lederman1984}), which tests the null hypothesis that the sSFR distributions of groups/clusters and of the respective field are drawn from the same parent distribution by measuring the maximum difference between the two cumulative distribution functions. The test returns the probability of the measured difference in the two cumulative distribution functions occurring if the two samples are drawn from the same parent distribution. A smaller $P$-value indicates that the two distributions are unlikely to be drawn from the same parent distribution. Table~\ref{tableSFRrisKS} lists the results of the K$-$S test, comparing the sSFR distributions of group/cluster galaxies in different radial bins with the sSFR distribution of the respective field galaxies (see Table~\ref{tableSFRnumbers} for the considered number of galaxies). The results for clusters are less significant compared with those for groups likely due to the smaller cluster sample. The group SFR distributions for the ranges $0\leq (R/R_{200}) <1$ and $1\leq (R/R_{200}) <2$ and the field sSFR distribution do not belong to the same parent population. For the comparison between group galaxies in $2\leq (R/R_{200})\leq 4.5$ and the field, the K$-$S test result is only marginally significant. Regarding clusters, the difference is significant for members in the radial bins $0\leq (R/R_{200}) <1$, marginal for $1\leq (R/R_{200}) <2$ and not significant for the range closest to the field $2\leq (R/R_{200})\leq 4.5$. This is in agreement with a scenario where the number of galaxies that have encountered the group/cluster core decreases beyond $R_{200}$, while the number of infallers and line-of-sight interlopers increases. This dilutes the population of low sSFR galaxies (as seen in Figures~\ref{SFRhistogramsgroups} and \ref{SFRhistogramsclusters}), and therefore the distribution becomes more field-like at large group/cluster-centric distances.

However, the K$-$S test does not probe differences in the tails of the distributions, but it is more sensitive to the behavior of the distributions close to their median values. Thus, following the procedure of \citet{Zabludoff1993} and \citet{Owers2009}, the sSFR distribution is approximated by a series of Gauss-Hermite functions up to order 4 and we estimate the strength of the asymmetric and symmetric departures from a Gaussian shape. Figures~\ref{SFRhistogramsgroups} and~\ref{SFRhistogramsclusters} show the $\log(\rm sSFR_{H\alpha})$ histograms for the group/cluster star-forming galaxies in different radial ranges and in the respective field reported in Table~\ref{tableSFRnumbers}, and they list the mean value ($<\log(\rm sSFR_{H\alpha})>$), standard deviation ($\sigma_{\log(\rm sSFR_{H\alpha})}$), skewness ($h_{3}$) and kurtosis ($h_{4}$) with the respective $P$-values showing the significance of these terms ($P[h_{3}]$ and $P[h_{4}]$). The group/cluster sSFR distributions are characterized by larger $\sigma_{\log(\rm sSFR_{H\alpha})}$ and lower $<\log(\rm sSFR_{H\alpha})>$ values compared to the associated field. Moreover, the group sSFR histograms in the radial ranges $0\leq (R/R_{200}) <1$ and $1\leq (R/R_{200}) <2$ show significant evidence for asymmetry, with $h_3=-0.054$ for both, meaning heavier tails for the lower sSFR side of the distribution relative to a Gaussian shape. This indicates a galaxy population with suppressed sSFR in groups. We do not find significant skewness values for the cluster sSFR distributions likely because of the smaller number of clusters and galaxies therein. 


\begin{table}
  \caption{Full Sample: star-forming galaxies in radial bins.}
  \label{tableSFRnumbers}
	\centering
\begin{tabular}{l c c r}
  \hline
  \noalign{\smallskip}
  \hline
  \noalign{\smallskip}
  \multicolumn{1}{c}{\small{}}
	&\multicolumn{1}{c}{\small{$(M_{200}/M_{\odot})$}}
	&\multicolumn{1}{c}{\small{$(R/R_{200})$}}
	&\multicolumn{1}{c}{\small{$N_{\rm gals}$}}
	\\
	\hline
        \noalign{\smallskip}
Groups  &$10^{12}-10^{14}$   & 0.0$-$1.0 & 1945\\
Groups  &$10^{12}-10^{14}$   & 1.0$-$2.0 &1068\\
Groups  &$10^{12}-10^{14}$  & 2.0$-$4.5 &  1677\\
Field   & $10^{12}-10^{14}$   &4.5$-$9.0&4177\\
 \noalign{\smallskip}
	\hline
        \noalign{\smallskip}       
Clusters&$10^{14}-10^{15}$ & 0.0$-$1.0 &446\\
Clusters&$10^{14}-10^{15}$& 1.0$-$2.0&273\\
Clusters&$10^{14}-10^{15}$ & 2.0$-$4.5& 752\\
        Field& $10^{14}-10^{15}$ & 4.5$-$9.0&2390\\
        \noalign{\smallskip}
	\hline
        \noalign{\smallskip}
\end{tabular}
\end{table}

\begin{table}
	\caption{K$-$S test.}
	\centering
        \label{tableSFRrisKS}
\begin{tabular}{l c r}
  \hline
  \noalign{\smallskip}
  \hline
  \noalign{\smallskip}
	\multicolumn{1}{l}{\small{$(M_{200}/M_{\odot})$}}
	&\multicolumn{1}{c}{\small{$(R/R_{200})$}}
	&\multicolumn{1}{c}{\small{$P$}}
	\\
	\hline
        \noalign{\smallskip}
$10^{12}-10^{14}$  & 0.0$-$1.0        &$6.76\times 10^{-16}$\\
$10^{12}-10^{14}$   & 1.0$-$2.0        &$7.36\times10^{-11}$\\
$10^{12}-10^{14}$  & 2.0$-$4.5        &$1.95\times10^{-2~\,}$\\
 $10^{12}-10^{14}$  & 0.0$-$4.5        &$5.72\times10^{-15}$ \\
 \noalign{\smallskip}
	\hline
        \noalign{\smallskip}       
$10^{14}-10^{15}$   & 0.0$-$1.0        &$1.47\times10^{-4}$\\
$10^{14}-10^{15}$   & 1.0$-$2.0        &$5.59\times10^{-2}$\\
$10^{14}-10^{15}$   & 2.0$-$4.5        &$7.79\times10^{-1}$\\
$10^{14}-10^{15}$   & 0.0$-$4.5        &$1.34\times10^{-2}$\\
        \noalign{\smallskip}
	\hline
        \noalign{\smallskip}
\end{tabular}
\tablecomments{$P$-values from comparing the group/cluster sSFR distribution binned in radius with that of the respective field.}
\end{table}


\subsection{SFR$-$galaxy stellar mass relationship}
Since the comparison between the sSFR distributions of star-forming group/cluster and field galaxies indicates that the median sSFRs are lower in groups/clusters than in the field, we compare the SFR$-M_{*}$ relationship for group and cluster members with that for the respective field in order to check whether and how this relation changes with environment. We also analyze whether the SFR quenching is stronger for low-mass galaxies compared with the high-mass ones, since this effect has been found by several studies (e.g., \citealp{vonderLinden2010, Ras2012,Davies2016a,Schaefer2016}).

\begin{figure*}
  \centering
 \includegraphics[width=0.87\textwidth]{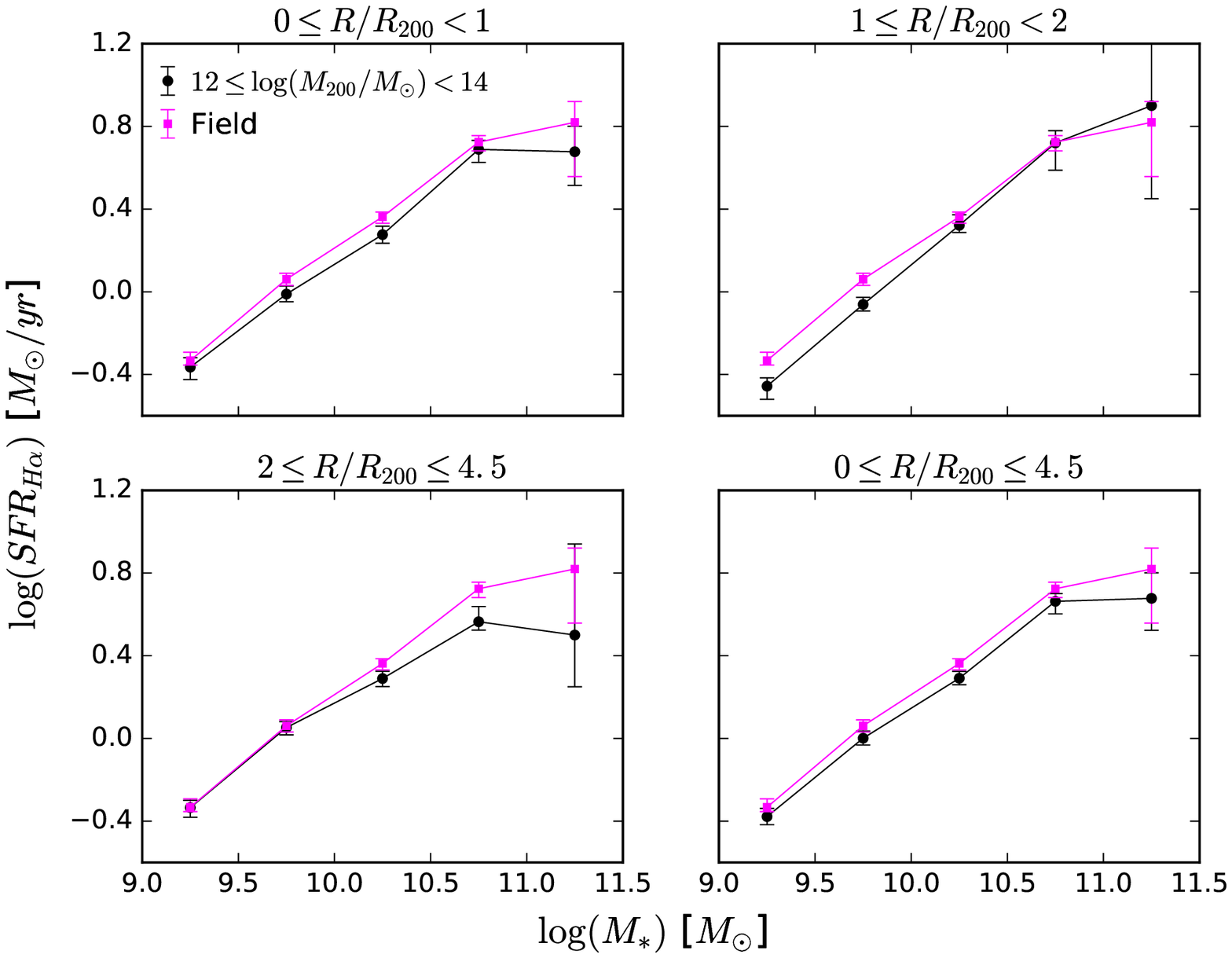}
 \includegraphics[width=0.87\textwidth]{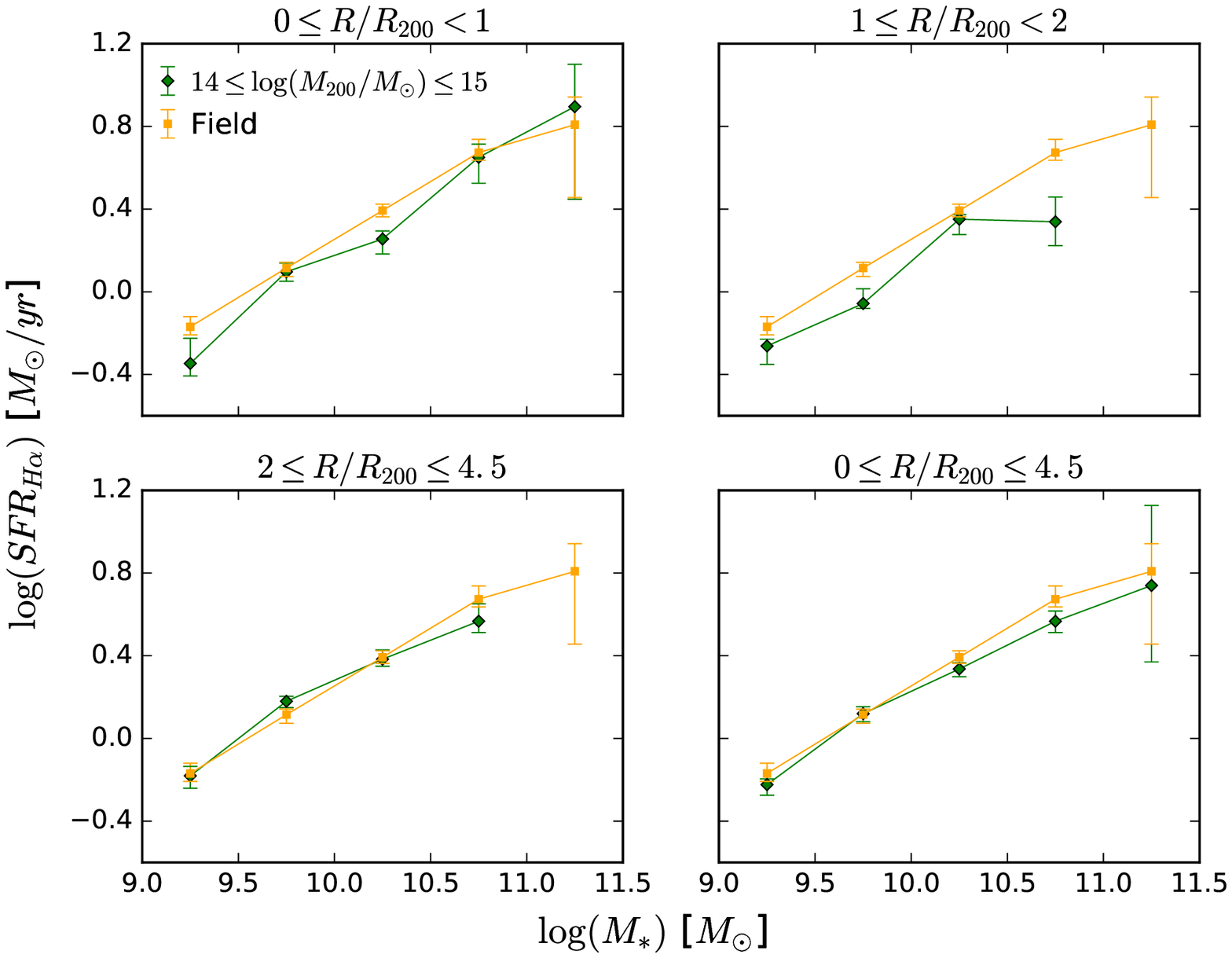}
 \caption{$\rm SFR_{H\alpha}$ as a function of $M_{*}$ for star-forming group/cluster members (black dots/green diamonds) in different radial bins and for the respective field galaxies (magenta/orange squares). We plot median values in 5 $M_{*}$ bins with errors at the 68\% c.l. At fixed galaxy stellar mass, star-forming members have lower $\rm SFR_{H\alpha}$ compared to field galaxies. This difference becomes less visible for radial ranges closer to the field.}
\label{SFRvsMstar}
\end{figure*}

Figure~\ref{SFRvsMstar} shows $\rm SFR_{H\alpha}$ as a function of $M_{*}$ for star-forming members of the Full Sample in different radial ranges out to $4.5\,R_{200}$, respectively of groups and clusters, and for the assigned field galaxies at $4.5< (R/R_{200})\leq 9$ according to the result of Figure~\ref{sSFRvsRadius}. The numbers of galaxies for each range in $M_{200}$ and $R_{200}$ are reported in Table~\ref{tableSFRnumbers}. We plot the median SFR$-M_{*}$ relations binning galaxies with $10^{9}\leq (M_{*}/M_{\odot})\leq 10^{11.5}$ in 5 bins, since the range $10^{11.5}< (M_{*}/M_{\odot})\leq 10^{12}$ is populated by too few galaxies. In all cases the SFR goes up as $M_{*}$ increases, forming the well known and often named star formation main sequence \citep{Daddi2007,Noeske2007}.

At fixed stellar mass, star-forming group/cluster members are shifted towards lower median values of $\rm SFR_{H\alpha}$ when compared with the values of the respective field galaxies. This means that there is a galaxy population with suppressed star formation activity in groups/clusters which is less noticeable in the field.

The difference between the median $\rm SFR_{H\alpha}$ of cluster members and field galaxies becomes less visible as the radial range is closer to the field and for the range $2\leq (R/R_{200})\leq 4.5$ there is no shift between the SFR$-M_{*}$ relationships in the cluster and field environments. This highlights the presence of an infalling star-forming population in clusters from the field. Our outcomes for clusters are in agreement with those shown in Figures 1 and 2 of \citet{Paccagnella2016}, who compared the SFR$-M_{*}$ relationship of star-forming galaxies in 31 OMEGAWINGS clusters at $0.04<z<0.07$ with that of the field, considering only galaxies with $M_{*}> 10^{9.8} M_{\odot}$. They found a population of quenched star-forming galaxies in these clusters that is rare in the field, suggesting that the transition from star-forming to passive occurs on a sufficiently long timescale to be observed. However, \citet{Paccagnella2016} observed a more evident transition galaxy population with respect to our result, likely due to the fact that their sample contains many more cluster galaxies than ours, i.e. 9242 and 1546 cluster galaxies in total, respectively.

We do not find a stronger suppression in SFR for low-mass galaxies in both groups and clusters, but the shift in SFR occurs over the whole range in $M_{*}$. \citet{Ras2012} found that the SFR suppression is strongest for low-mass galaxies with $M_{*}\leq 10^{9} M_{\odot}$, while the decline is negligible for high-mass galaxies with $M_{*}> 10^{10} M_{\odot}$. They detected a dependence of the sSFR$-$radius relation on $M_{*}$ that we do not observed in our data. However, we do not probe galaxies with $M_{*}\leq 10^{9} M_{\odot}$ as \citet{Ras2012}, and a future inclusion of these galaxies could be crucial for shedding light on this effect.

\begin{figure*}
  \centering
 \includegraphics[width=0.87\textwidth]{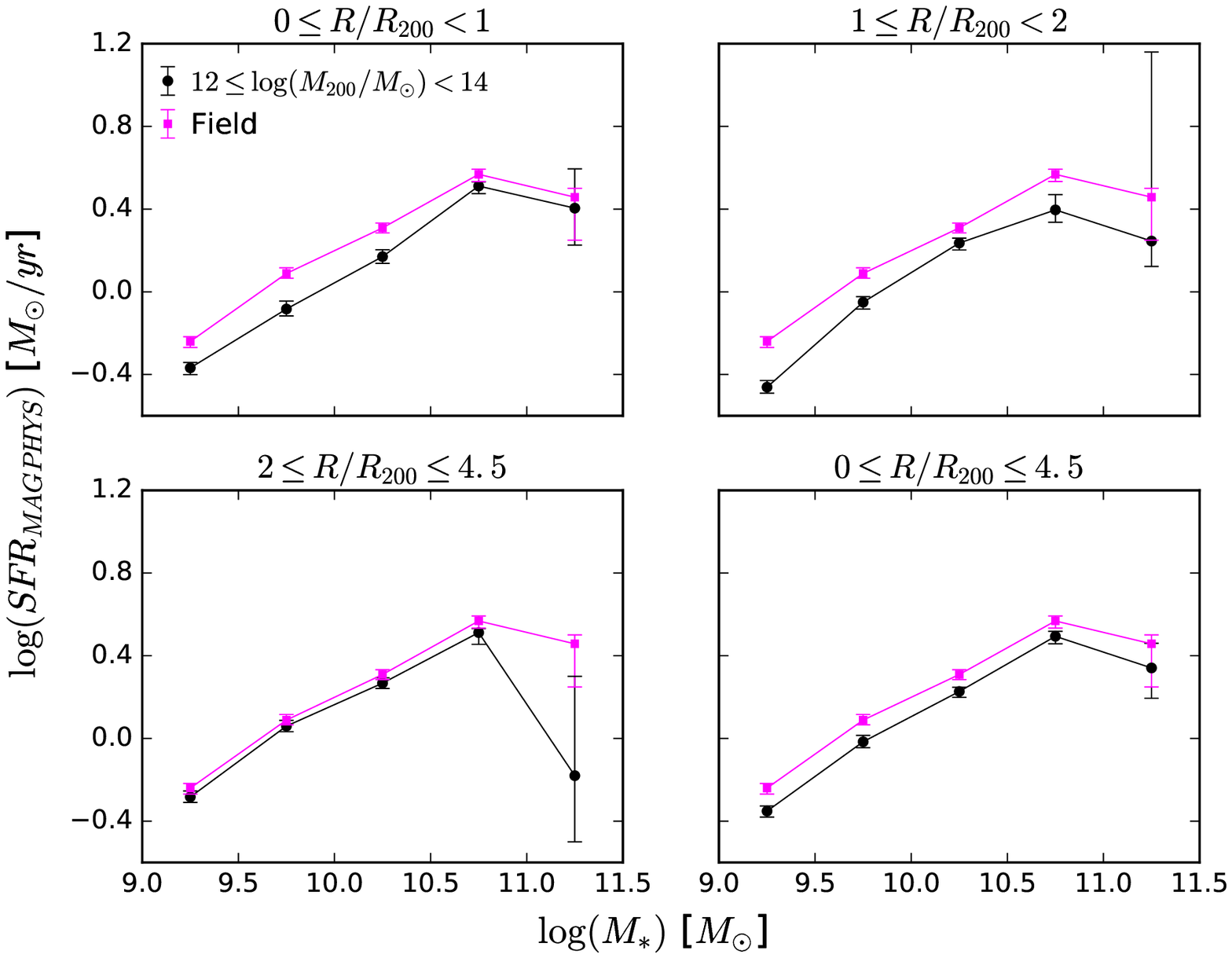}
 \includegraphics[width=0.87\textwidth]{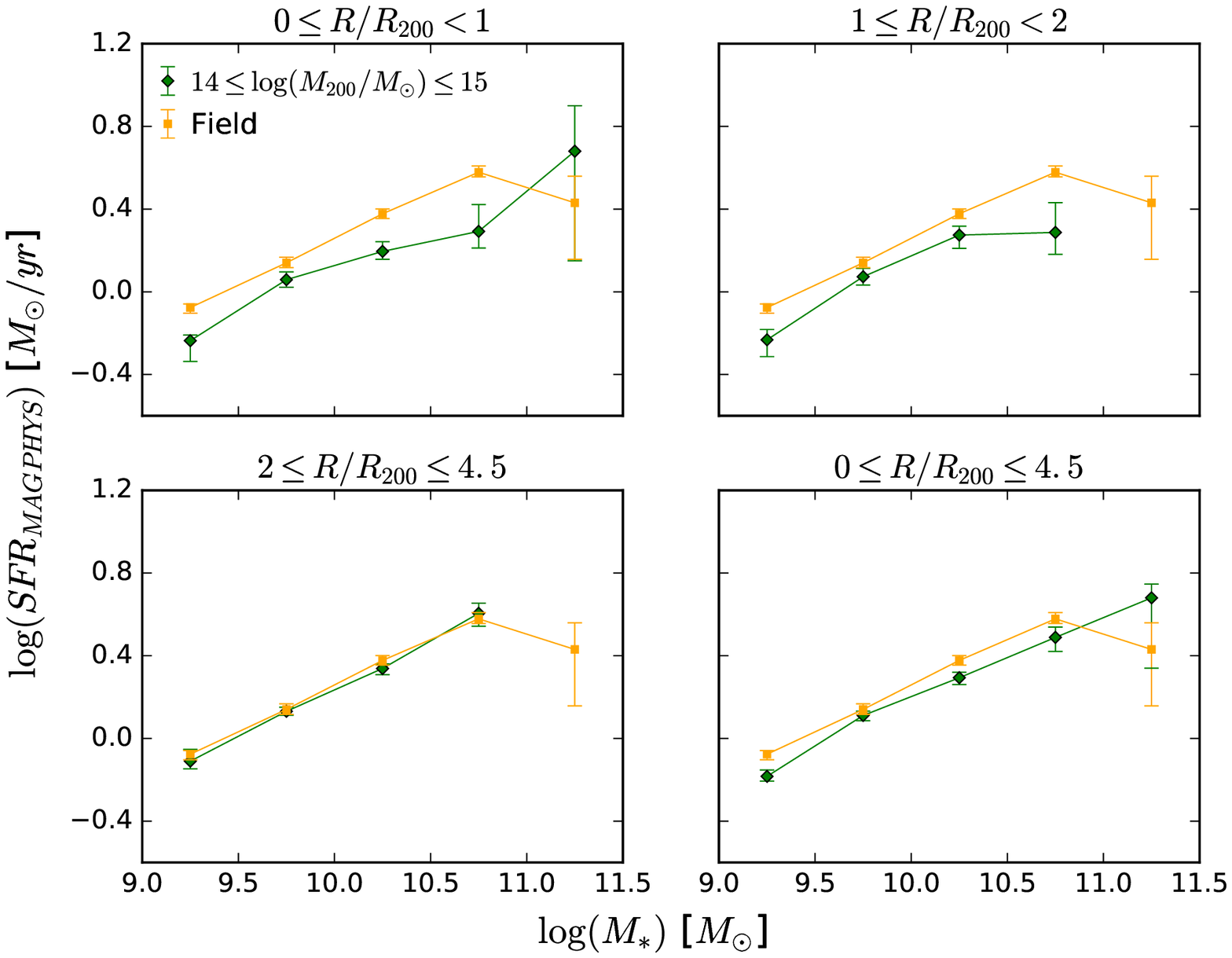}
 \caption{$\rm SFR_{MAGPHYS}$ as a function of $M_{*}$ for star-forming group/cluster members (black dots/green diamonds) in different radial bins and for the respective field galaxies (magenta/orange squares). We plot median values in 5 $M_{*}$ bins with errors at the 68\% c.l. At fixed galaxy stellar mass, star-forming members have lower $\rm SFR_{MAGPHYS}$ compared to field galaxies. This difference becomes less visible for radial ranges closer to the field.}
	\label{photoSFRvsMstar}
\end{figure*}

Finally, we consider the photometric estimators of the star formation activity, plotting median values of $\rm SFR_{MAGPHYS}$ in $M_{*}$ bins, in order to compare these results with those obtained using $\rm SFR_{H\alpha}$. We use star-forming galaxies with available $\rm SFR_{MAGPHYS}$ in the same radial and halo mass ranges as in the previous case and in the respective field. Figures~\ref{photoSFRvsMstar} shows that there is a change of the SFR$-M_{*}$ relation with the environment, i.e. groups versus field and clusters versus field, respectively. At fixed galaxy stellar mass, group galaxies are characterized by lower values of SFR compared to field galaxies. This result agrees with that found in Figure~\ref{SFRvsMstar} and confirms that the star-forming galaxies in groups have lower SFRs than those in the field. The strongest difference in SFR between the benchmark field sample and the group galaxies occurs in the smallest radius bin and then the shift becomes less marked with increasing radius. The same finding is observed in both Figures~\ref{SFRvsMstar} and~\ref{photoSFRvsMstar} for cluster members and field galaxies. As in the case of $\rm SFR_{H\alpha}$, we do not observe a stronger SFR quenching for low-mass galaxies in groups and clusters. In conclusion, we observe the same outcomes using $\rm SFR_{H\alpha}$ or $\rm SFR_{MAGPHYS}$.

\section{Discussion} \label{discussion}
We investigate the distributions of passive and star-forming galaxies in radial space, projected phase space and velocity space (Sections 3.1$-$3.3, respectively). The analysis of the radial space confirms that the inner regions of groups/clusters are mainly populated by passive galaxies, whereas the outskirts are dominated by star-forming galaxies. This finding is in agreement with many previous works in both groups and clusters (e.g., \citealp{Postman1984, Carlberg2001, Lewis2002, Girardi2003, Gomez2003, vonderLinden2010,Wilman2012}). The study of the group/cluster PPS reveals that the passive and star-forming populations inhabit different regions. Star-forming galaxies generally inhabit regions which simulations have shown to be dominated by infalling galaxies, while passive galaxies inhabit regions of the PPS dominated by the virialized populations. \citet{Mahajan2011} observed this segregation of galaxy populations for a nearby SDSS cluster sample, \citet{Oman2013} and \citet{Oman2016} for a sample made by both groups and clusters, but we find the same result for a sample composed only by low-mass halos. Finally, the galaxy type segregation in velocity space is well investigated for cluster galaxies (e.g., \citealp{Biviano1997,Haines2015,Barsanti2016}), but it is less studied in the group environment. We confirm for groups the kinematic segregation of galaxy populations also observed by \citet{Lares2004} for 2dFGRS low-mass halos. We find galaxy stellar mass segregation in velocity space: more massive galaxies are slower down and this effect is stronger for passive galaxies compared with the star-forming population. Since the most massive galaxies are thought to be the most luminous ones, these results are in agreement with those of \citet{Biviano1992} and \citet{Girardi2003} who found luminosity segregation in velocity space for the brightest passive galaxies in nearby clusters and groups, respectively. We note that \citet{Kafle2016} observed no galaxy stellar mass segregation with radius for the GAMA group galaxies, while we detect this effect in velocity space. These segregations in the radial, PPS and velocity spaces strongly suggest that star-forming galaxies have been recently accreted onto groups from the surrounding field, while the passive population have resided in the group for a much longer period of time. Moreover, the galaxy stellar mass segregation suggests a picture in which the dynamical friction process slows the motion of the most massive passive galaxies.


For star-forming galaxies we observe a decline in sSFR with decreasing projected group-centric (Section 3.4) out to $9\,R_{200}$ in agreement with the outcome of \citet{Ras2012}, while the absence of this sSFR-radius relationship within $1.5\,R_{200}$ agrees with the conclusions of \citet{Ziparo2013}. This suggests that the conflicting results observed by \citet{Ras2012} and \citet{Ziparo2013} are due to the different radial range selected, i.e. out to $\sim10\,R_{200}$ and $\sim1.5\,R_{200}$, respectively. The observation of SFR suppression at large group-centric radius and not within the region most directly affected by the group environment ($<1.5\,R_{200}$) suggests that the star formation in group galaxies is stopped slowly. As a consequence, the quenching timescale probably is of the order of few Gyr and comparable to the group crossing time.

The comparison of the sSFR distributions of star-forming group/cluster galaxies with that of the respective field (Section 3.5) shows that they do not belong to the same parent population and that there is a shift towards lower median sSFR values for the group/cluster distributions. The sSFR distributions for the inner regions of groups have a significant skewness, indicating heavier tails for the lower sSFR side of the distribution relative to a Gaussian shape and highlighting the presence of a galaxy population with suppressed sSFR in groups.

The analysis of the SFR$-M_{*}$ relation (Section 3.6) suggests that there is a population of quenched star-forming galaxies in groups but not in the field, meaning that the transition from star-forming to quenched occurs on a sufficiently long timescale to be observed. This is in agreement with the conclusions of \citet{Paccagnella2016}, who found a population of quenched star-forming galaxies in clusters and not in the field, implying a long timescale for the quenching process. These galaxies are mainly observed within the central regions of the halos, while the SFR$-M_{*}$ relation in groups becomes more consistent with the field for the outer regions. 

Finally, \citet{Ras2012} found a dependence of the star formation$-$radius relation on $M_{*}$ for groups, observing a more pronounced suppression for galaxies with $M_{*}\leq 10^{9} M_{\odot}$. However, we do not probe galaxies with such low stellar mass and we are unable to test the SFR suppression for these low-mass galaxies.

\section{Summary and conclusions} \label{conclusions}
We study 1197 GAMA halos with $10^{12}\leq (M_{200}/M_{\odot})\leq 10^{15}$ at $0.05\leq z\leq 0.2$ and we select member galaxies with $10^{9}\leq (M_{*}/M_{\odot})\leq 10^{12}$ (the Full Sample). We consider two different ranges in $M_{200}$ in order to analyze the SFR distributions and the PPS of galaxies in low- and high-mass halos, i.e groups with $M_{200}/M_{\odot}=10^{12}-10^{14}$ and clusters with $M_{200}/M_{\odot}=10^{14}-10^{15}$. We divide galaxies into different spectral types, i.e. passive, star-forming and AGN/composite. We restrict the analysis of the passive and star-forming fractions to galaxies with $10^{10}\leq (M_{*}/M_{\odot})\leq 10^{12}$ in halos at $0.05\leq z\leq 0.15$ to ensure completeness $>$95\% (the Restricted Sample).

We study the passive and star-forming galaxy populations of the Restricted Sample in the radial and PPS spaces. We also investigate the velocity space for the Full Sample, obtaining the following results:

\begin{enumerate}
\item The fraction of passive galaxies decreases by a factor of $\sim2$ from the group center towards 2.5$\,R_{200}$, whereas the fraction of the star-forming galaxies goes up by a factor of $\sim1.5$ within the same radial distance.
  
\item The virialized region in the PPS is dominated  by passive galaxies, while the fraction of star-forming galaxies is much higher in the infalling region.
  
\item Passive and star-forming galaxies are segregated in velocity space with the velocity profile of star-forming members higher than that of the passive ones, according to the $\chi^{2}-$test. 
  
\item The most massive passive galaxies are segregated in velocity according to the Spearman test. 
\end{enumerate}

The analysis of the star-forming galaxies of the Full Sample in the group/cluster environment leads to important outcomes which can be summarized as follows:

\begin{enumerate}
\item The SFR of star-forming member galaxies declines towards the halo inner regions. The sSFR decreases by a factor of $\sim 1.2$ from the field to the halo center, and the decline starts in the range $2.5\leq (R/R_{200})\leq 4.5$. Considering only the region within $1.5\,R_{200}$, we do not detect any change of the sSFR with radius.

\item The sSFR distributions of star-forming members and field galaxies do not belong to the same parent population, according to the K$-$S test. The group sSFR distributions for the inner radial regions show a significant evidence for asymmetry, indicating heavier tails for the lower sSFR side of the distributions relative to a Gaussian shape.
  
\item  At fixed galaxy stellar mass, star-forming members have lower SFRs compared to field galaxies.
  
\item The SFR$-$radius and SFR$-M_{*}$ relationships agree using both H$\alpha$ emission lines and the SED-fitting code MAGPHYS as SFR estimators.
\end{enumerate}

These results suggest that the star-forming group galaxies are recently accreted and represent an infalling population from the field to the halo. The decline in SFR of star-forming galaxies with radius, coupled with their status as recent infallers, implies that the group environment quenches the star formation of the galaxies as they fall into the halo.

In conclusion, our analysis for groups reveals that many results observed for clusters are also visible in these lower-mass halos, i.e. suppression of SFR with decreasing radius, galaxy populations with low SFR, passive versus star-forming$-$radius relation and PPS, kinematic and galaxy stellar mass segregations.

\acknowledgements
We thank the referee for the useful comments. MSO acknowledges the funding support from the Australian Research Council through a Future Fellowship (FT140100255). SB acknowledges funding support from the Australian Research Council through a Future Fellowship (FT140101166). GAMA is a joint European–Australasian project based around a spectroscopic campaign using the Anglo-Australian Telescope. The
GAMA input catalog is based on data taken from the SDSS and
the UKIRT Infrared Deep Sky Survey. Complementary imaging of
the GAMA regions is being obtained by a number of independent
survey programs including GALEX MIS, VST KiDS, VISTA
VIKING, WISE, Herschel-ATLAS, GMRT and ASKAP providing
UV to radio coverage. GAMA is funded by the STFC (UK), the
ARC (Australia), the AAO, and the participating institutions. The
GAMA website is http://www.gama-survey.org/.

\bibliographystyle{aasjournal}
\bibliography{biblio2}



\end{document}